\documentclass[aps,prd,preprintnumbers,groupedaddress,showpacs,superscriptaddress,manuscript,floatfix]{revtex4}
\usepackage{epsfig}
\usepackage{latexsym}
\usepackage{graphicx}
\usepackage{bm}
\usepackage{longtable}
\usepackage{color}
\usepackage{ulem}
\usepackage{amssymb,amsmath}

\newcommand{\kslash}{k\kern-1ex /}
\newcommand{\pslash}{p\kern-1ex /}
\newcommand{\qslash}{q\kern-1ex /}
\newcommand{\lslash}{l\kern-1ex /}
\newcommand{\sslash}{s\kern-1ex /}
\newcommand{\Dslash}{D\kern-1.2ex /}

\newcommand{\beqa}{\begin{eqnarray}}
\newcommand{\eeqa}{\end{eqnarray}}

\newcommand{\bd}{\begin{description}}
\newcommand{\ed}{\end{description}}

\newcommand{\ben}{\begin{eqnarray}}
\newcommand{\een}{\end{eqnarray}}

\def\lsim{\raise0.3ex\hbox{$<$\kern-0.75em\raise-1.1ex\hbox{$\sim$}}}
\def\gsim{\raise0.3ex\hbox{$>$\kern-0.75em\raise-1.1ex\hbox{$\sim$}}}
\def\simgt{\rlap{\lower 3.5 pt\hbox{$\mathchar \sim$}}\raise 1.0pt \hbox {$>$}}
\def\simlt{\rlap{\lower 3.5 pt\hbox{$\mathchar \sim$}}\raise 1.0pt \hbox {$<$}}

\begin{document}

\bibliographystyle{apsrev}

\preprint{UTCCS-P-63, UTHEP-628}

\title{Two-Nucleon Bound States in Quenched Lattice QCD}

\author{T.~Yamazaki}
\affiliation{Kobayashi-Maskawa Institute for the Origin of Particles and the Universe, Nagoya University, Naogya, Aichi 464-8602, Japan}
\affiliation{Center for Computational Sciences, University of Tsukuba,
Tsukuba, Ibaraki 305-8577, Japan}
\author{Y.~Kuramashi}
\affiliation{Center for Computational Sciences, University of Tsukuba,
Tsukuba, Ibaraki 305-8577, Japan}
\affiliation{Graduate School of Pure and Applied Sciences,
University of Tsukuba, Tsukuba, Ibaraki 305-8571, Japan}
\affiliation{RIKEN Advanced Institute for Computational Science,
Kobe, Hyogo 650-0047, Japan}
\author{A.~Ukawa}
\affiliation{Center for Computational Sciences, University of Tsukuba,
Tsukuba, Ibaraki 305-8577, Japan}
\collaboration{PACS-CS Collaboration}

\date{
\today
}

\begin{abstract}
We address the issue of bound state in the two-nucleon system in lattice QCD. 
Our study is made in the quenched approximation at 
the lattice spacing of $a =0.128$ fm
with a heavy quark mass corresponding to $m_\pi = 0.8$ GeV.
To distinguish a bound state from an attractive scattering state,
we investigate the volume dependence of the energy difference 
between the ground state and the free two-nucleon state by changing the
spatial extent of the lattice from 3.1 fm to 12.3 fm.
A finite energy difference left in the infinite spatial volume limit
leads us to the conclusion that the measured ground states for 
not only spin triplet but also singlet channels
are bounded. Furthermore the existence of the bound state is confirmed by
investigating the properties of the energy for the first excited state
obtained by 2$\times$2 diagonalization method.
The scattering lengths for both channels are evaluated by
applying the finite volume formula derived by L\"uscher to the energy of the first excited states.
\end{abstract}

\maketitle

\section{Introduction}
\label{sec:introduction}

The strong interaction dynamically generates
a  hierarchical structure: three quarks are bound to form a nucleon with an energy of 1~GeV,  
and nucleons are in turn bound to form nuclei with a binding energy of 10~MeV or so per nucleon. 
This is a multi-scale physics that computational physics should explore, and lattice QCD is responsible 
for explaining the nature of nuclei based on first principles.

Recently, the present authors have made a first attempt to directly construct 
the Helium-3 and Helium-4 nuclei 
from quarks and gluons in lattice QCD.  
In order to control statistical errors in the Monte Carlo evaluation of the Helium Green's function
as well as quark contractions whose number  factorially increases with the nuclear mass number, 
calculations were carried out at a rather heavy degenerate 
up and down quark mass corresponding to
$m_\pi=0.8$ GeV in quenched QCD~\cite{Yamazaki:2009ua}. 
We successfully confirmed the formation of Helium nuclei as a bound state.  The key 
was a systematic change of the spatial size of the lattice over a sufficiently wide range that 
allowed a reliable extrapolation  to the infinite volume limit.
After our finding of the Helium nuclei, NPLQCD Collaboration reported an 
evidence of the H di-baryon bound state in $N_f = 2 + 1$ QCD 
at $m_\pi = 0.39$ GeV investigating
the volume dependence of the energy shift
from twice of the $\Lambda$ baryon mass~\cite{Beane:2009py,Beane:2010hg}.
This was  followed by HALQCD Collaboration which also presented
an evidence of the H di-baryon, but in degenerate $N_f = 3$ QCD
at $m_\pi = 0.67$--1.02 GeV based on analysis with the effective potential
measured by the two-baryon wave function~\cite{Inoue:2010es}. 

The situation is markedly  different for deuteron.  This is the simplest nucleus composed of 
two nucleons in the spin triplet channel, and yet evidence based on lattice QCD for bound state has never been reported. 
It is already quite some time ago that  a first analysis of the two-nucleon system was made 
in quenched QCD~\cite{{Fukugita:1994na},{Fukugita:1994ve}}.
Much more recently, studies were made with a partially-quenched 
mixed action~\cite{Beane:2006mx}
and $N_f = 2+1$ anisotropic Wilson action~\cite{Beane:2009py}.
Extraction of the potential between two nucleons has been investigated in
quenched and 2+1 flavor QCD~\cite{{Ishii:2006ec},{Aoki:2009ji},{Aoki:2008hh}}. 
All these studies, however, tried to calculate the two-nucleon scattering
lengths assuming, based primarily on model considerations with nuclear potentials, 
 that the deuteron becomes unbound for 
the heavy quark mass, corresponding to $m_\pi \simgt \ 0.3$ GeV, 
employed in their simulations.

It is time to check the validity of this assumption.
We need to investigate whether the bound state exists or not in the
heavy quark mass region, where studies  so far have been carried out, using the arsenal of 
methods solely within lattice QCD.  
If there is a bound state, the ground state energy never yield
the scattering length if substituted into the L\"uscher's finite volume 
formula~\cite{Luscher:1986pf,Luscher:1990ux}.
In such a  case, the scattering length should be obtained from 
the energy of the first excited state.

We carry out two types of calculations at a heavy quark mass 
corresponding to $m_\pi = 0.8$ GeV in quenched QCD.
The first one is a conventional analysis  in which  we
investigate the volume dependence of the energy shift for the ground state.
Different volume dependence is expected for scattering 
and bound states.
In the second one we investigate the energy level of the first excited state
employing the diagonalization method~\cite{Luscher:1990ck}
to separate the  first excited state from the ground state near the threshold of 2$m_N$.
If we find the ground state slightly below the threshold and the first excited state slightly above it,
then such a configuration of the two lowest levels is consistent with the ground state 
being a bound state and the first excited state a scattering state with almost zero relative 
momentum.  
This method was previously used in a scalar QED simulation
to distinguish a system with or without a bound state~\cite{Sasaki:2006jn}.

Hereafter we call the analyses employed in the first and second calculations 
the single state and two state analyses, respectively.
We also refer to the configuration sets used in the two calculations
as the first and second ensembles.
We should note that the $^3$S$_1$-$^3$D$_1$ mixing is neglected in this paper, 
since we restrict ourselves to measure states 
in the small relative momentum region.

This paper is organized as follows.
Section~\ref{sec:1st} presents the results of the single state analysis
in the first calculation together with the simulation details.
In Sec.~\ref{sec:2nd} we explain the operators
employed in the diagonalization method   
and examine the results obtained by the two state analysis.
Conclusions and discussions are summarized in Sec.~\ref{sec:summary}.

\section{Single state analysis}
\label{sec:1st}

Let us first present the results of the single state analysis
for the $^3$S$_1$ and $^1$S$_0$ channels.

\subsection{Simulation details}

The first ensemble is exactly the same 
as in the previous work of Ref.~\cite{Yamazaki:2009ua}.
We explain the parameters once again for clarity.

We generate quenched configurations with the 
Iwasaki gauge action~\cite{Iwasaki:1983cj} at $\beta = 2.416$ whose
lattice spacing is $a=0.128$ fm, corresponding to $a^{-1} = 1.541$ GeV,
determined with $r_0=0.49$ fm 
as an input~\cite{AliKhan:2001tx}.
We employ the HMC algorithm with the 
Omelyan-Mryglod-Folk integrator~\cite{{Omelyan:2003om},{Takaishi:2005tz}}.
The step size is
chosen to yield reasonable acceptance rate presented 
in Table~\ref{tab:conf_meas}.
We take three lattice sizes, 
$L^3\times T = 24^3 \times 64$, $48^3 \times 48$ and $96^3 \times 48$, 
to investigate the spatial volume dependence of the energy 
difference between the two-nucleon ground state and twice  the nucleon mass.
The physical spatial extents are 3.1, 6.1 and 12.3 fm, respectively.

We use the tadpole improved Wilson action 
with $c_{\mathrm{SW}} = 1.378$~\cite{AliKhan:2001tx}.
Since it becomes harder to obtain a reasonable signal-to-noise ratio at
lighter quark masses for the multi-nucleon system, 
we employ a heavy quark mass at $\kappa = 0.13482$ which gives
$m_\pi = 0.8$ GeV for the pion mass and $m_N = 1.6$ GeV for the nucleon mass. 
Statistics is increased by repeating the measurement of 
the correlation functions
with the source points in different time slices on each configuration.
The number of configurations and measurements on each configuration
are listed in Table~\ref{tab:conf_meas}.
We separate successive measurements by 100 trajectories 
with $\tau=1$ for the trajectory length.
The errors are estimated by jackknife analysis choosing 200 
trajectories for the bin size.

The quark propagators are solved with
the periodic boundary condition 
in all the spatial and temporal directions
using the exponentially smeared source
\begin{equation}
q^\prime(\vec{x},t) = \sum_{\vec{y}} A\, e^{-B|{\vec x} - \vec{y}|} q(\vec{y},t)
\label{eq:smear}
\end{equation}
after the Coulomb gauge fixing.
On each volume we employ two sets of   
smearing parameters: $(A,B) = (0.5,0.5)$, $(0.5,0.1)$ 
for $L=24$ and $(0.5,0.5)$, $(1.0,0.4)$ for $L=48$ and 96.
The onset of ground state can be confirmed by 
consistency of effective masses with different sources as shown later. 
Hereafter the nucleon operators using the first and the second smearing 
parameter sets are referred to as ${\cal O}_{1}$ and ${\cal O}_{2}$, 
respectively.

The interpolating operator for the proton is defined as 
\begin{equation}
p_\alpha = \varepsilon_{abc}([u_a]^tC\gamma_5 d_b)u_c^\alpha,
\label{eq:def:proton}
\end{equation}
where $C = \gamma_4 \gamma_2$ and $\alpha$ and $a,b,c$ are the Dirac index and
the color indices, respectively.
The neutron operator $n_\alpha$ is obtained 
by replacing $u_c^\alpha$ by $d_c^\alpha$ 
in the proton operator.
To save the computational cost
we use the nonrelativistic quark operator, in which the Dirac index
is restricted to the upper two components.

The two-nucleon operators for the $^3$S$_1$ and $^1$S$_0$ 
channels are given by 
\begin{eqnarray}
NN_{^3{\mathrm S}_1}(t) &=& \frac{1}{\sqrt{2}}\left[
p_+(t) n_+(t) - n_+(t) p_+(t)
\right],
\label{eq:def:3S1}\\
NN_{^1{\mathrm S}_0}(t) &=& 
\frac{1}{\sqrt{2}}\left[
p_+(t) p_-(t) - p_-(t) p_+(t)
\right].
\label{eq:def:1S0}
\end{eqnarray}
For the source operator we insert the {\it smeared} quark fields of 
Eq.~(\ref{eq:smear}) for each nucleon operator located at the same spatial point $\vec{x}$.
Each nucleon in the sink operator, on the other hand,  is composed of the {\it point} quark fields,  
and projected to have zero spatial
momentum. We call this type of sink operator the {\it point sink} operator.
In the spin triplet channel
the operators for other two spin components are constructed in a similar way.
We increase the statistics by averaging over the three spin components.

\subsection{Numerical results}

\begin{figure}[!t]
\includegraphics*[angle=0,width=0.6\textwidth]{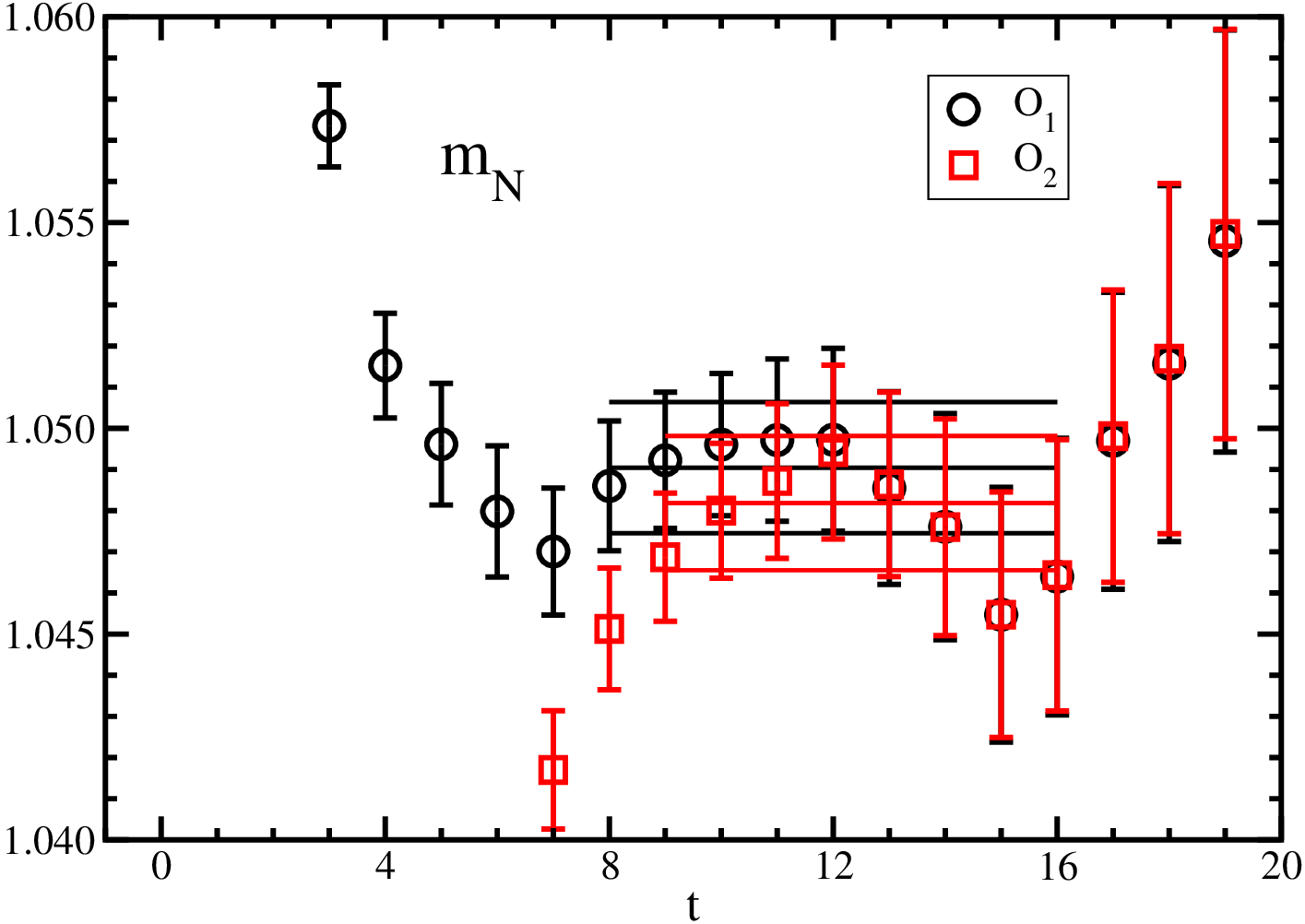}
\caption{
Effective mass of nucleon
with ${\cal O}_1$ (circle) and ${\cal O}_2$ (square) sources on 
(6.1 fm)$^3$ box in lattice unites.
Fit results with one standard deviation error band are expressed
by solid lines.
\label{fig:eff_n_1st}
}
\end{figure}

Let us first present the effective mass of the nucleon 
on the (6.1 fm)$^3$ box in Fig.~\ref{fig:eff_n_1st}.
We observe that the signals with the ${\cal O}_{1,2}$ source operators
are clean 
and the plateaux show reasonable consistency with each other.
The exponential fit results with one standard deviation errors
are denoted by the solid lines.
They also show the consistency between
the results from the two nucleon correlation functions.

\begin{figure}[!t]
\includegraphics*[angle=0,width=0.6\textwidth]{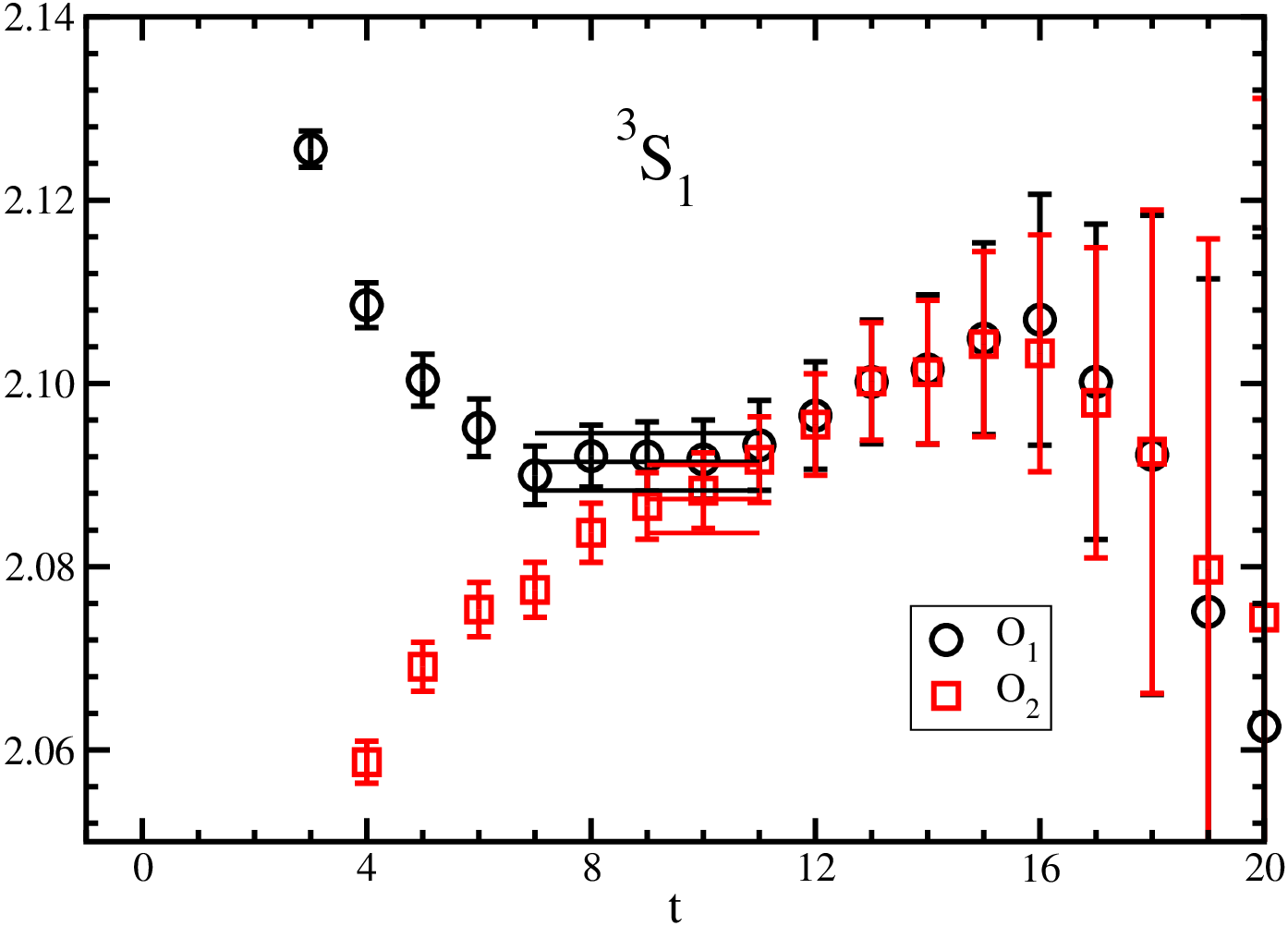}
\caption{
Effective energy of two-nucleon for $^3$S$_1$ channel
with ${\cal O}_1$ (circle) and ${\cal O}_2$ (square) sources on
(6.1 fm)$^3$ box in lattice unites.
Fit results with one standard deviation error band are expressed
by solid lines.
\label{fig:eff_3S1}
}
\end{figure}

\begin{figure}[!t]
\includegraphics*[angle=0,width=0.6\textwidth]{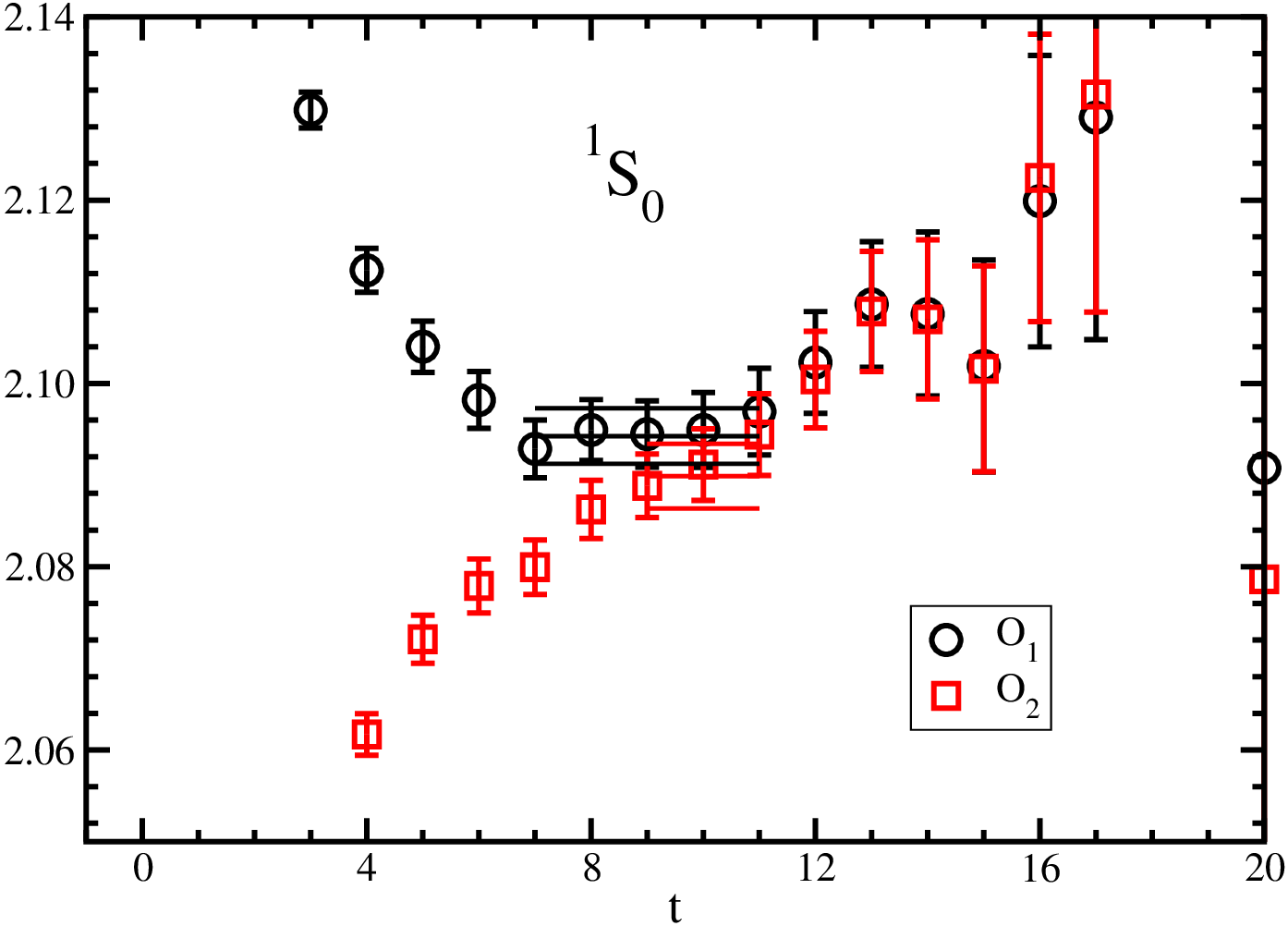}
\caption{
Same as Fig.~\ref{fig:eff_3S1} for $^1$S$_0$ channel.
\label{fig:eff_1S0}
}
\end{figure}

Figure~\ref{fig:eff_3S1} shows the effective energy plots for 
the two-nucleon correlation functions 
with the ${\cal O}_{1,2}$ operators 
in the $^3$S$_1$ channel on the same volume
as in the above.
We find clear signals up to $t\approx 12$, beyond which
statistical fluctuation dominates.
The effective masses with the different sources show a reasonable
agreement in the plateau region.
The result of  exponential fit over the plateau region is presented by the solid lines for each 
operator in the figure.
Similar behavior of the effective energy is observed in
the $^1$S$_0$ channel as shown in Fig.~\ref{fig:eff_1S0}.

In order to determine the energy shift $\Delta E_L=E_{NN}-2E_N$ 
precisely in each volume, 
we define the ratio of the two-nucleon correlation function divided by
the nucleon correlation function squared,
\begin{equation}
R(t) = \frac{G_{NN}(t)}{\left(G_N(t)\right)^2},
\end{equation}
where the same source operator is chosen for $G_{\mathrm{NN}}(t)$ 
and $G_N(t)$.
The effective energy shift is extracted as
\begin{equation}
\Delta E_L^{\mathrm{eff}} = \ln \left(\frac{R(t)}{R(t+1)}\right).
\label{eq:delE_L}
\end{equation}

\begin{figure}[!t]
\includegraphics*[angle=0,width=0.6\textwidth]{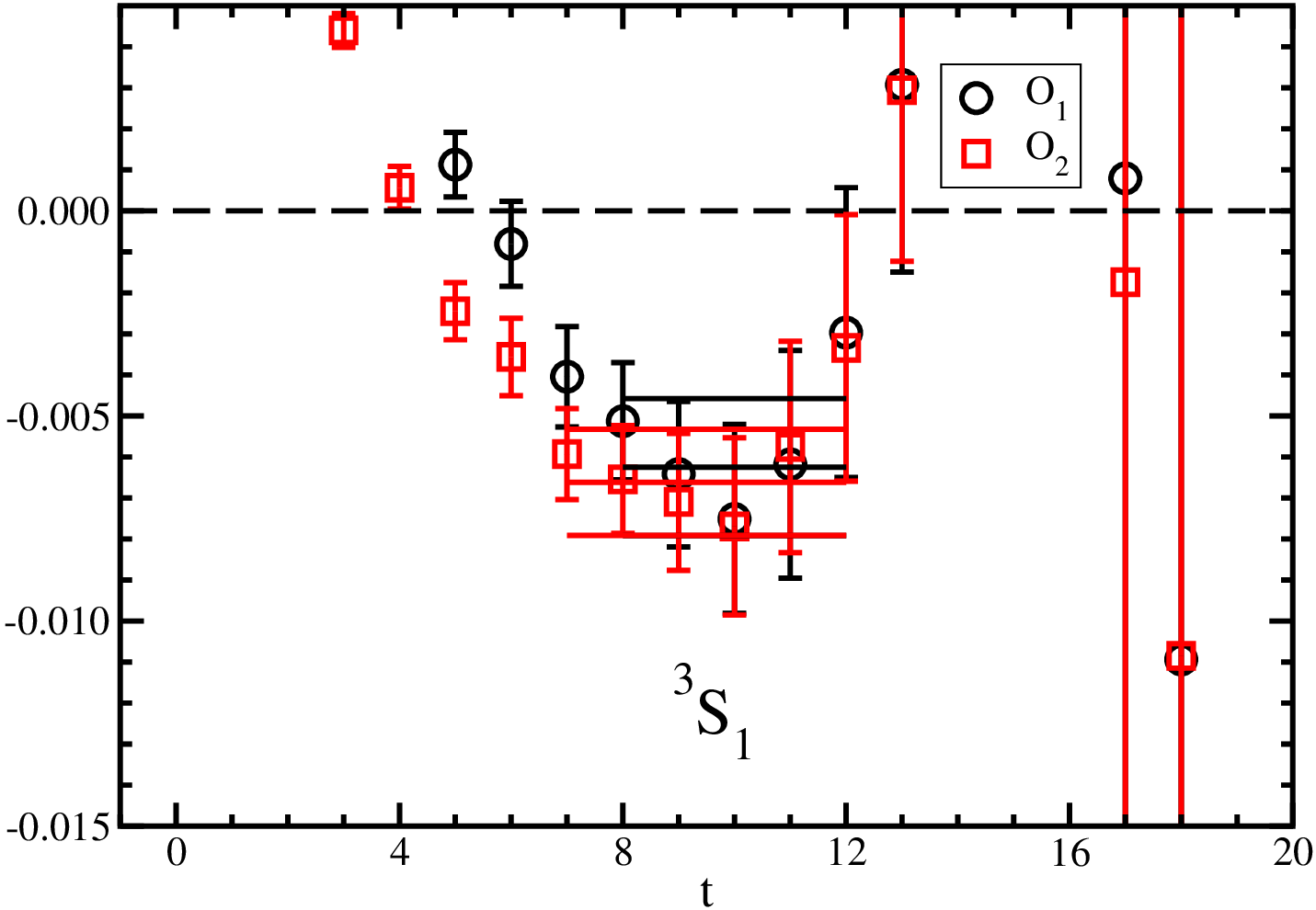}
\caption{
Effective energy shift of two-nucleon 
$\Delta E_L^{\mathrm{eff}}$ for $^3$S$_1$ channel
with ${\cal O}_1$ (circle) and ${\cal O}_2$ (square) sources 
on (6.1 fm)$^3$ box in lattice units.
Fit results with one standard deviation error band are expressed 
by solid lines.
\label{fig:eff_R_3S1}
}
\end{figure}

\begin{figure}[!t]
\includegraphics*[angle=0,width=0.6\textwidth]{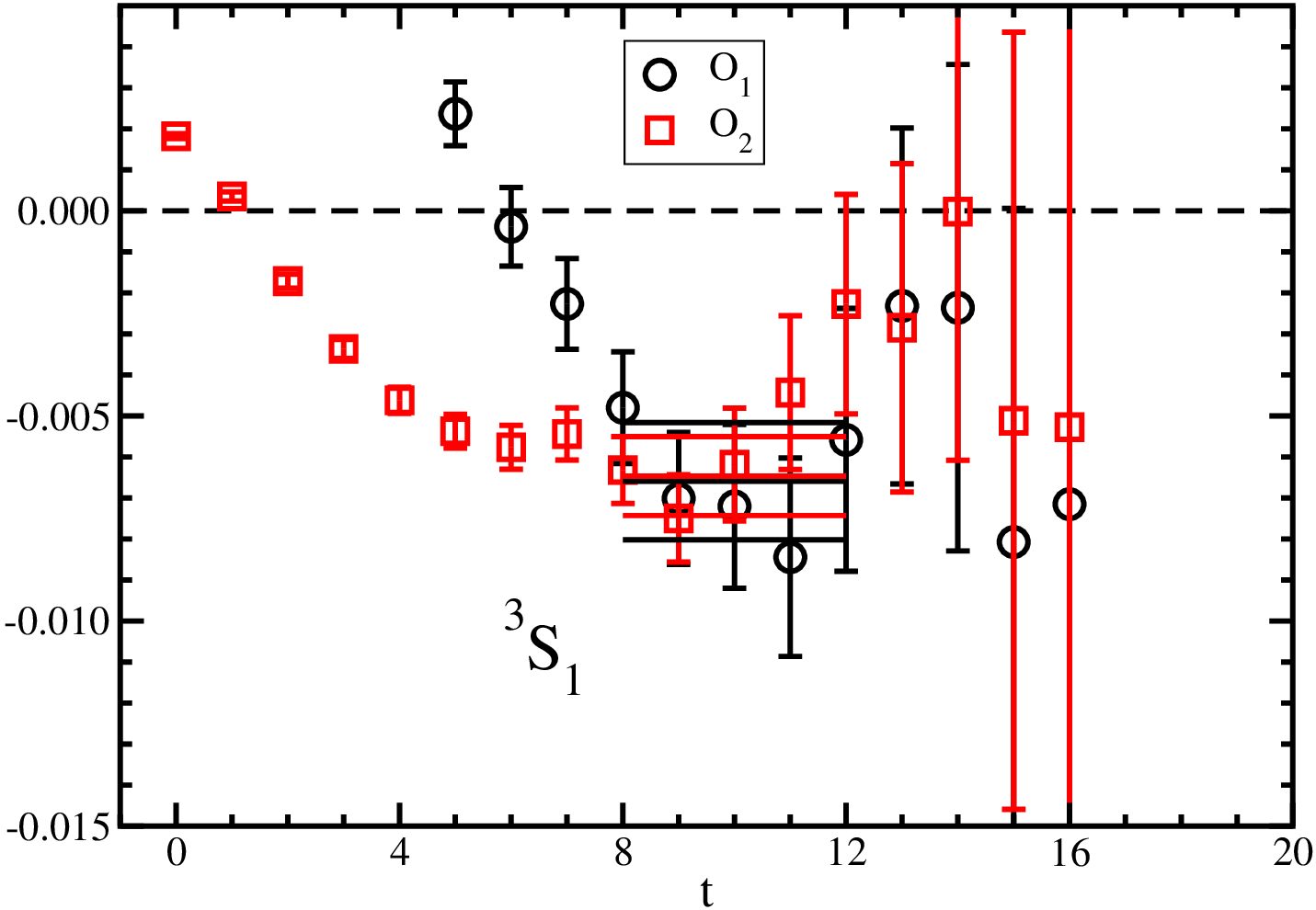}
\caption{
Same as Fig.~\ref{fig:eff_R_3S1} on (3.1 fm)$^3$ box.
\label{fig:eff_R_3S1_24}
}
\end{figure}

\begin{figure}[!t]
\includegraphics*[angle=0,width=0.6\textwidth]{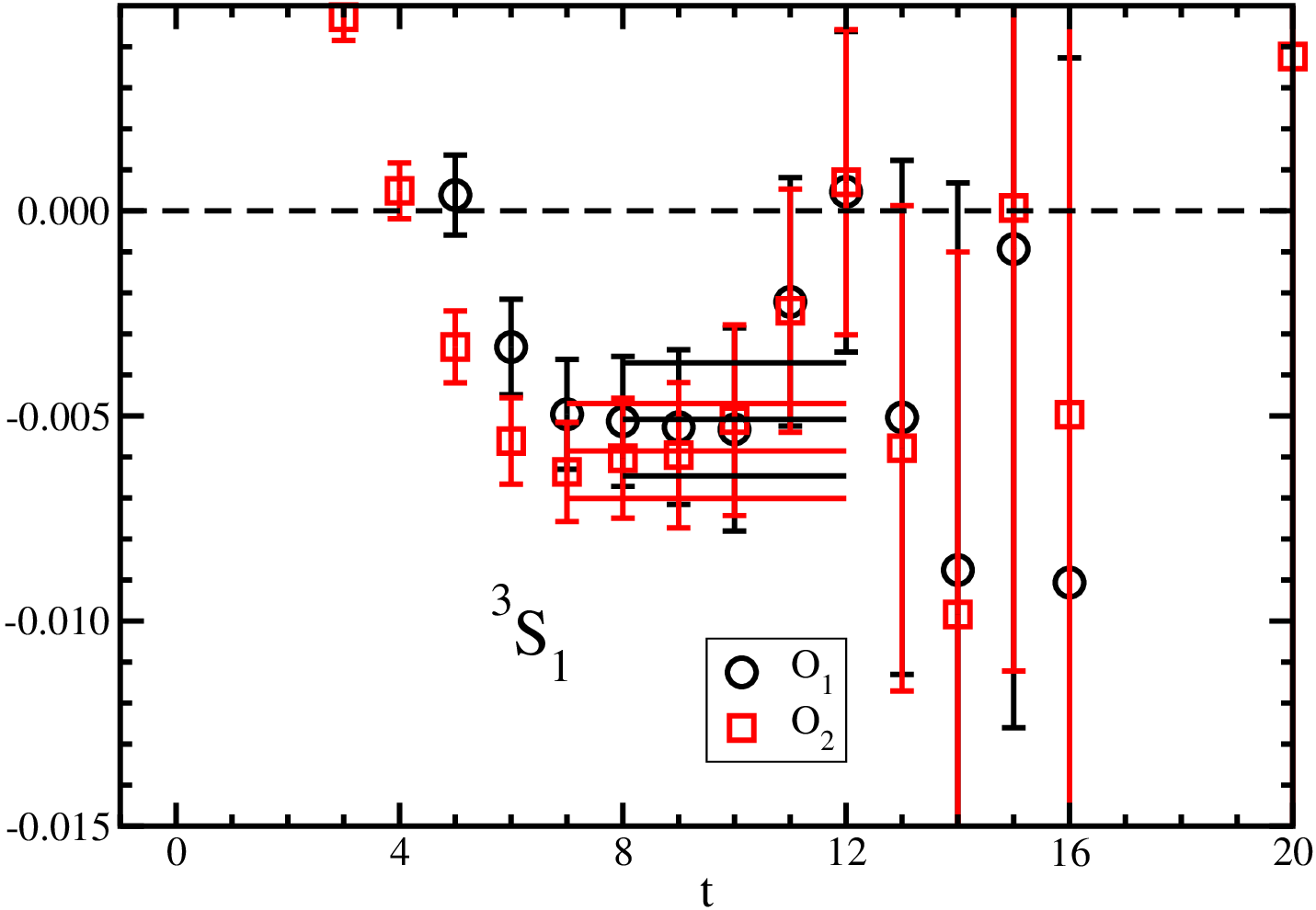}
\caption{
Same as Fig.~\ref{fig:eff_R_3S1} on (12.3 fm)$^3$ box.
\label{fig:eff_R_3S1_96}
}
\end{figure}

In Fig.~\ref{fig:eff_R_3S1} we present typical results of 
time dependence of $\Delta E_L^{\mathrm{eff}}$ for 
the ${\cal O}_{1,2}$ sources in the $^3$S$_1$ channel, both of which
show negative values 
beyond the error bars in the plateau region of $t=8$--11.
Note that this plateau region is reasonably consistent 
with that for the effective mass
of the two-nucleon correlation functions in Fig.~\ref{fig:eff_3S1}.
The signals of $\Delta E_L^{\mathrm{eff}}$ 
are lost beyond $t\approx 12$ because of 
the large fluctuations in the two-nucleon correlation functions.
We determine $\Delta E_L$ by an exponential fit of the ratio in 
the plateau region, $t=8$--12 for ${\cal O}_1$ and 
$t=7$--12 for ${\cal O}_2$, respectively.
The  systematic error of the fit 
is estimated from the difference of the central values of the fit results with
the minimum or maximum time slice changed by $\pm 1$.
We obtain a similar quality for the signal 
for different boxes of (3.1 fm)$^3$ and (12.3 fm)$^3$
as shown in Figs.~\ref{fig:eff_R_3S1_24} and \ref{fig:eff_R_3S1_96}, 
respectively. 

\begin{figure}[!t]
\includegraphics*[angle=0,width=0.6\textwidth]{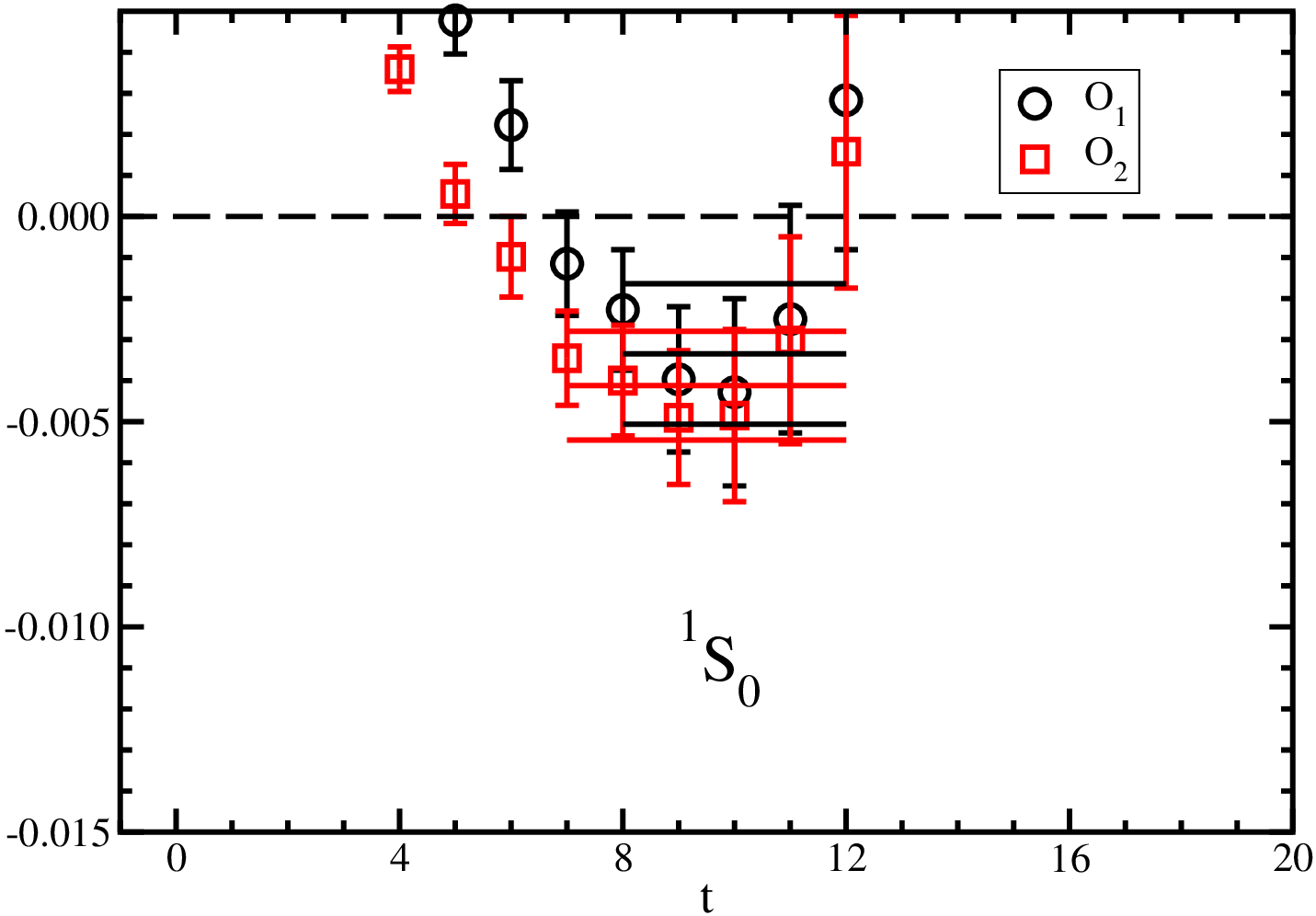}
\caption{
Same as Fig.~\ref{fig:eff_R_3S1} for $^1$S$_0$ channel.
\label{fig:eff_R_1S0}
}
\end{figure}

The result for the $^1$S$_0$ channel on the (6.1 fm)$^3$ box
is shown in Fig.~\ref{fig:eff_R_1S0}.
We find that the effective energy shift $\Delta E_L^{\mathrm{eff}}$ 
is negative beyond the error bars,
though its absolute value is smaller than the $^3$S$_1$ case.
The energy shift $\Delta E_L$ is determined in the same way 
as for the $^3$S$_1$ channel.

\begin{figure}[!t]
\includegraphics*[angle=0,width=0.6\textwidth]{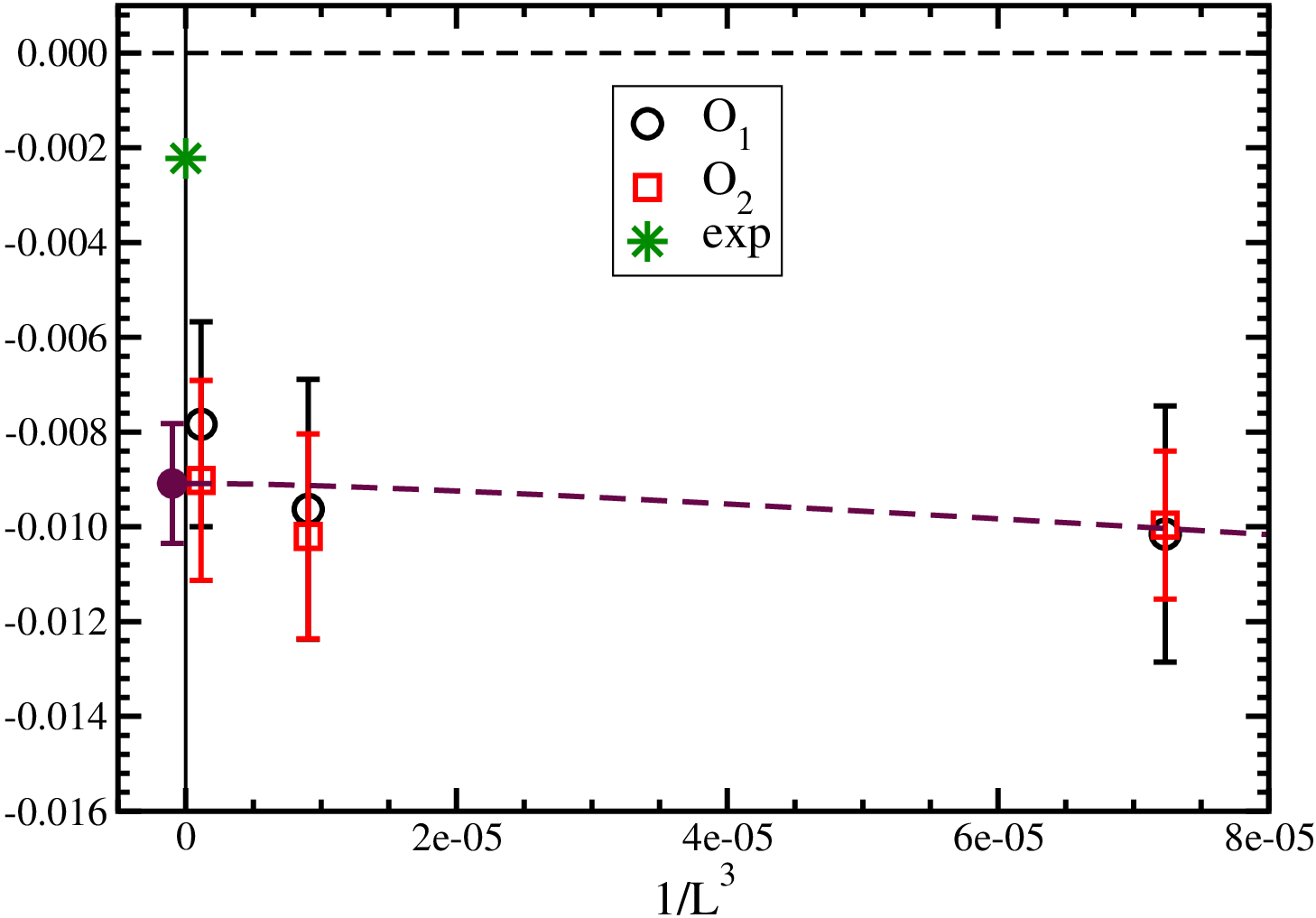}
\caption{
Spatial volume dependence of $\Delta E_L = E_{NN} - 2 m_N$ in GeV units
for $^3$S$_1$ channel with ${\cal O}_1$ (circle) 
and ${\cal O}_2$ (square) sources.
Statistical and systematic errors are added in quadrature.
Extrapolated results to the infinite spatial 
volume limit (filled circle) 
and experimental values (star) are also presented.
\label{fig:dE_3S1_wo}
}
\end{figure}

The volume dependence of the energy shift $\Delta E_L$ 
for the $^3$S$_1$ channel
is plotted as a function of $1/L^3$ in Fig.~\ref{fig:dE_3S1_wo}. 
Table~\ref{tab:dE} summarizes the numerical values
of $\Delta E_L$ on three spatial volumes, where the statistical
and systematic errors are presented in the first and second parentheses,
respectively.
The results for the ${\cal O}_{1,2}$ sources are consistent 
within the error bars.
Little volume dependence for $\Delta E_L$ indicates 
a bound state, rather than the $1/L^3$ dependence expected for a 
scattering state, for the ground state in the $^3$S$_1$ channel.

The binding energy in the infinite spatial volume limit in Table~\ref{tab:dE} 
is extracted by a simultaneous fit of the data for the ${\cal O}_{1,2}$ sources 
employing 
the fit function including a finite volume effect for the 
two-particle bound state~\cite{Beane:2003da,Sasaki:2006jn},
\begin{equation}
\Delta E_L = -\frac{\gamma^2}{m_N}\left\{
1 + \frac{C_\gamma}{\gamma L} \sum^{\hspace{6mm}\prime}_{\vec{n}}
\frac{\exp(-\gamma L \sqrt{\vec{n}^2})}{\sqrt{\vec{n}^2}}
\right\},
\end{equation}
where $\gamma$ and $C_\gamma$ are free parameters, 
$\vec{n}$ is three-dimensional integer vector, 
and $\sum^\prime_{\vec{n}}$ denotes the summation without $|\vec{n}|=0$.
The binding energy, $-\Delta E_\infty$, is determined from $\gamma$,
\begin{equation}
-\Delta E_\infty = -\frac{\gamma^2}{m_N},
\end{equation}
where we assume
\begin{equation}
2\sqrt{m_N^2 - \gamma^2} - 2 m_N \approx -\frac{\gamma^2}{m_N}.
\end{equation}
The systematic error is estimated from the difference of the central values
of the fit results choosing different fit ranges
in the determination of $\Delta E_L$, and also using a constant fit as an alternative fit 
form.   Adding the statistical and systematic errors by quadrature, 
we obtain $-\Delta E_\infty$=9.1(1.3) MeV for the binding energy. 

We conclude that the ground state in the $^3$S$_1$ channel
is a bound state.  The provisos, of course, are that pion mass is quite heavy and that quark 
vacuum polarizations are left out.  Whether these are the reasons why the binding energy 
is about four times larger than the experimental value, 2.22 MeV, is an interesting issue for 
future study with lighter pion mass in full QCD.

\begin{figure}[!t]
\includegraphics*[angle=0,width=0.6\textwidth]{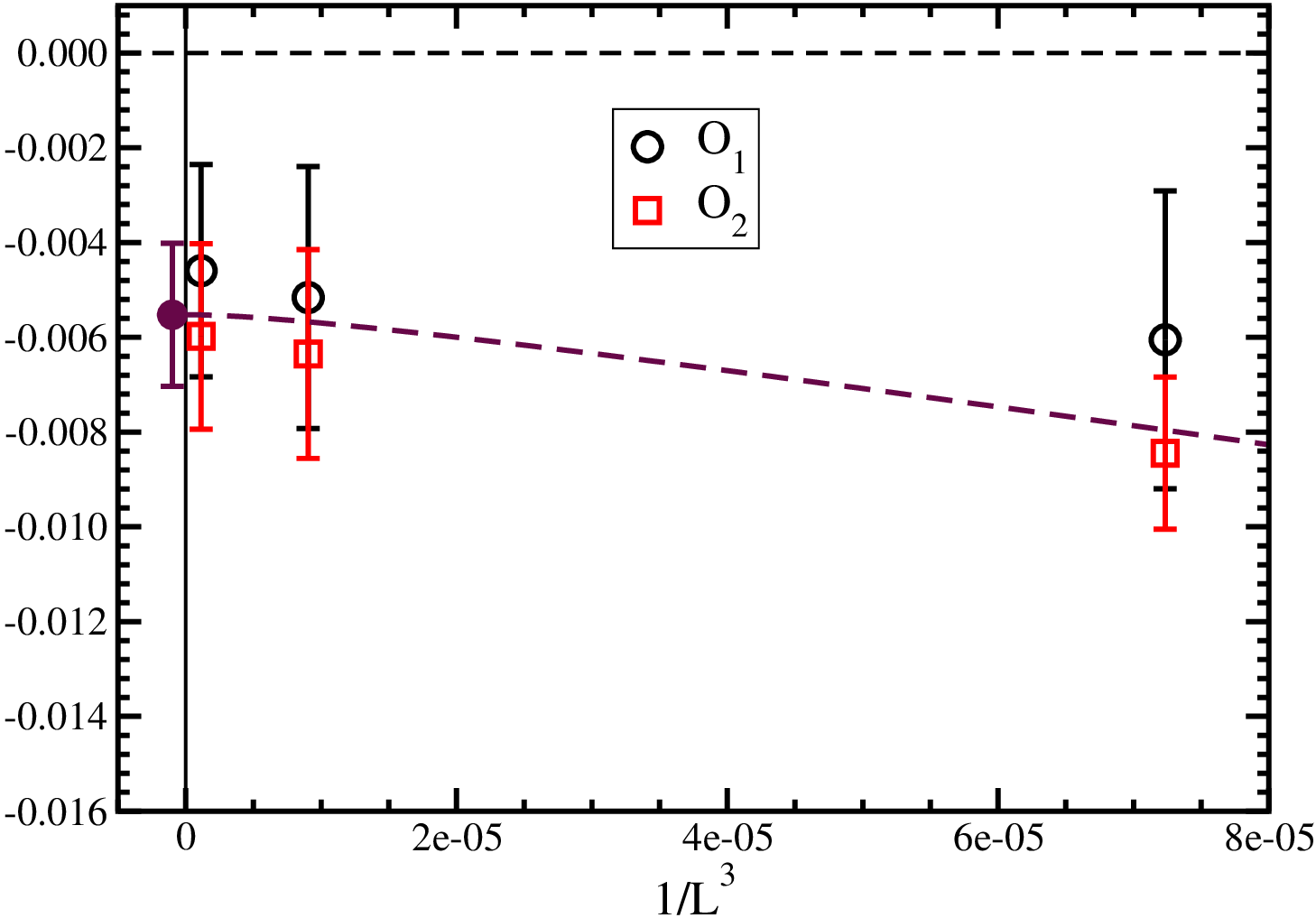}
\caption{
Same as Fig.~\ref{fig:dE_3S1_wo} for $^1$S$_0$ channel.
\label{fig:dE_1S0_wo}
}
\end{figure}

Figure~\ref{fig:dE_1S0_wo} plots the volume dependence 
of the energy shift $\Delta E_L$ 
for the $^1$S$_0$ channel, whose 
numerical values are summarized in Table~\ref{tab:dE}.
Employing the same analysis as in the $^3$S$_1$ channel,
we find that  $-\Delta E_\infty = 5.5(1.5)$ MeV 
in the infinite volume limit,
which is 3.7 $\sigma$ away from zero.
This tells us that the ground state in the $^1$S$_0$ channel
is also bound at $m_\pi = 0.8$ MeV.
Since the existence of the bound state in this channel is not expected
at the physical quark mass, it might be 
a consequence of much heavier quark mass used in our calculation.
Although there are several 
model calculations varying the up and down quark masses, they
are restricted around the physical 
values~\cite{{Epelbaum:2002gb},{Beane:2002xf},{Flambaum:2007mj},{Chen:2010yt}}.
It is an intriguing subject to check 
if the bound state in the $^1$S$_0$ channel 
disappears at lighter quark masses.
This is beyond the scope of this paper, however.

\section{Two-state analysis}
\label{sec:2nd}

In this section we present the results of analysis with the diagonalization 
method~\cite{Luscher:1990ck}.
The focus of the analysis is the characteristic feature, well known from quantum mechanics, 
that the existence of a bound state 
implies a scattering state just above the two particle threshold, and hence a negative 
scattering length.  
Our investigation is carried out with the diagonalization 
of 2$\times$2 correlation function matrix.

\subsection{Simulation details}

We work with two spatial extents,  4.1 fm and 6.1 fm. 
The corresponding lattice sizes are $L^3\times T = 32^3\times 48$
and $48^3 \times 48$, respectively.
The latter is the same size as in the first ensemble, but 
we regenerate independent configurations employing the same algorithm.
Most of the simulation parameters, including the gauge and fermion
actions, lattice spacing, quark mass, are identical to those
explained in Sec.~\ref{sec:1st}.  However, the number of configurations and the 
separation of trajectories between each measurement,
and the number of measurements on each configuration are different.
These numbers are tabulated in Table~\ref{tab:conf_meas_2nd}.
The errors are estimated by the jackknife analysis choosing 400 and
200 trajectories for the bin size on the (4.1 fm)$^3$ and (6.1 fm)$^3$ boxes, 
respectively.
These bin sizes are sufficiently large to remove the autocorrelation.
We use the same operators for the nucleons
and two-nucleons as in Eqs.(\ref{eq:def:proton}), (\ref{eq:def:3S1})  
and (\ref{eq:def:1S0}) choosing the nonrelativistic components.

The diagonalization method for the 2$\times$2 matrix requires 
two operators each at source and sink time slice, which are
explained in the following subsections.

\subsubsection{Source operators}

We use the two-nucleon operator composed of the
${\cal O}_1$ nucleon operator explained 
in Sec.~\ref{sec:1st} as one of
the source operators for diagonalization; from the single state analysis, 
we expect that it has good overlap with the ground state
of the two-nucleon system.
To reduce the statistical error as much as possible
we carry out more than one hundred measurements
on each configuration by changing the center of the smearing source 
in the spatial and temporal directions.

The diagonalization procedure requires another operator 
which reasonably overlaps to the first excited state.  If we envisage this 
to be a scattering state of the two nucleons with almost zero relative momentum, 
then a  possible candidate is an operator consisting of two nucleons each projected 
to zero spatial momentum.
Constructing such an operator at the source time slice can be done using $Z(3)$ 
noises for the quark fields.  It is empirically known, however, that 
statistical noise overwhelms signal in the (multi-)nucleon correlation function 
if the noise is spread over the entire spatial volume. 

The large fluctuation can be reduced by restricting noise to a subset of lattice sites  
at a fixed separation of $N_{\mathrm{mod}}$ in each spatial dimension 
at the source time slice.
This source operator, which we shall call the ${\cal O}_r$ source, is
defined by 
\begin{equation}
{\cal O}_r(t) = \frac{1}{N_{\mathrm{rand}}}\sum_{j=1}^{N_{\mathrm{rand}}}
\left[\sum_{\vec{x}\in V^\prime} \xi_j(\vec{x}) q^\prime(\vec{x},t)\right]^3,
\end{equation}
where $N_{\mathrm{rand}}$ is the number of the noise, and
the color and Dirac indices of the quark field are omitted for simplicity.
We use the smeared quark fields $q^\prime(\vec{x},t)$ of Eq.~(\ref{eq:smear}) 
after Coulomb gauge fixing, whose parameters are the same as 
in the ${\cal O}_1$ source
$(A,B)=(0.5,0.5)$ to obtain faster plateau of the nucleon state.
The smeared quark field is located at 
\begin{equation}
V^\prime = \left\{
\vec{x} = \vec{x}_0 + \vec{n} N_{\mathrm{mod}},\ ({\vec x})_i < L
\right\}
\end{equation}
with $\vec{x}_0$ being a reference position, 
$(\vec{x}_0)_i < N_{\mathrm{mod}}$,
and $\vec{n}$ being three-dimensional integer vector.
The complex $Z(3)$ random number $\xi_j(\vec{x})$ satisfies 
$\left(\xi_j(\vec{x})\right)^3 = 1$ and has the property
\begin{equation}
\lim_{N_{\mathrm{rand}}\to \infty}
\frac{1}{N_{\mathrm{rand}}}\sum_{j=1}^{N_{\mathrm{rand}}} 
\xi_j(\vec{x})\xi_j(\vec{y})\xi_j(\vec{z})
= \delta_{\vec{x},\vec{y}}\delta_{\vec{x},\vec{z}}.
\end{equation}
The parameters $N_{\mathrm{mod}}$ and $N_{\mathrm{rand}}$ 
for each calculation
are summarized in Table~\ref{tab:conf_meas_2nd}.

\subsubsection{Sink operators}

We also need two operators on the sink side to carry out diagonalization.  
Our idea is to employ the solution of 
the Helmholtz equation in three dimensions for the smearing function
of the two-nucleon sink operator,
\begin{equation}
W_{q^2}({\vec r}) = C_{q^2} \sum_{\vec n}\frac{e^{i(2\pi/L){{\vec n}\cdot {\vec r}}}}
{\vec{n}^2-q^2},
\label{eq:W_q}
\end{equation}
where $q^2$ is a  parameter,
$\vec{r}$ is the relative coordinate between two nucleons,
and $\vec{n}$ is three-dimensional integer vector.
The overall factor $C_{q^2}$ is determined from the normalization condition 
$|W_{q^2}(\vec{r}_{\mathrm{max}})| = 1$.
A similar calculation using the solution of 
the Helmholtz equation was previously reported in Ref.~\cite{Gockeler:1994rx}.

In the region of $|\vec{r}|$ closer to the origin, the smearing function 
should be modified from 
the free form to take into account the two-particle interaction.
One way is to calculate the two-particle wave function, as has been done for
the two-pion system~\cite{{Aoki:2005uf},{Sasaki:2008sv}} and 
the two-nucleon systems~\cite{{Ishii:2006ec},{Aoki:2009ji},{Aoki:2008hh}}, and use it 
as input.  We take a simpler alternative of modifying the smearing function by hand such that 
it behaves as a smooth constant function around the origin rather than a sharp increase or 
decrease which occurs for the free form. 

\begin{figure}[!t]
\includegraphics*[angle=0,width=0.6\textwidth]{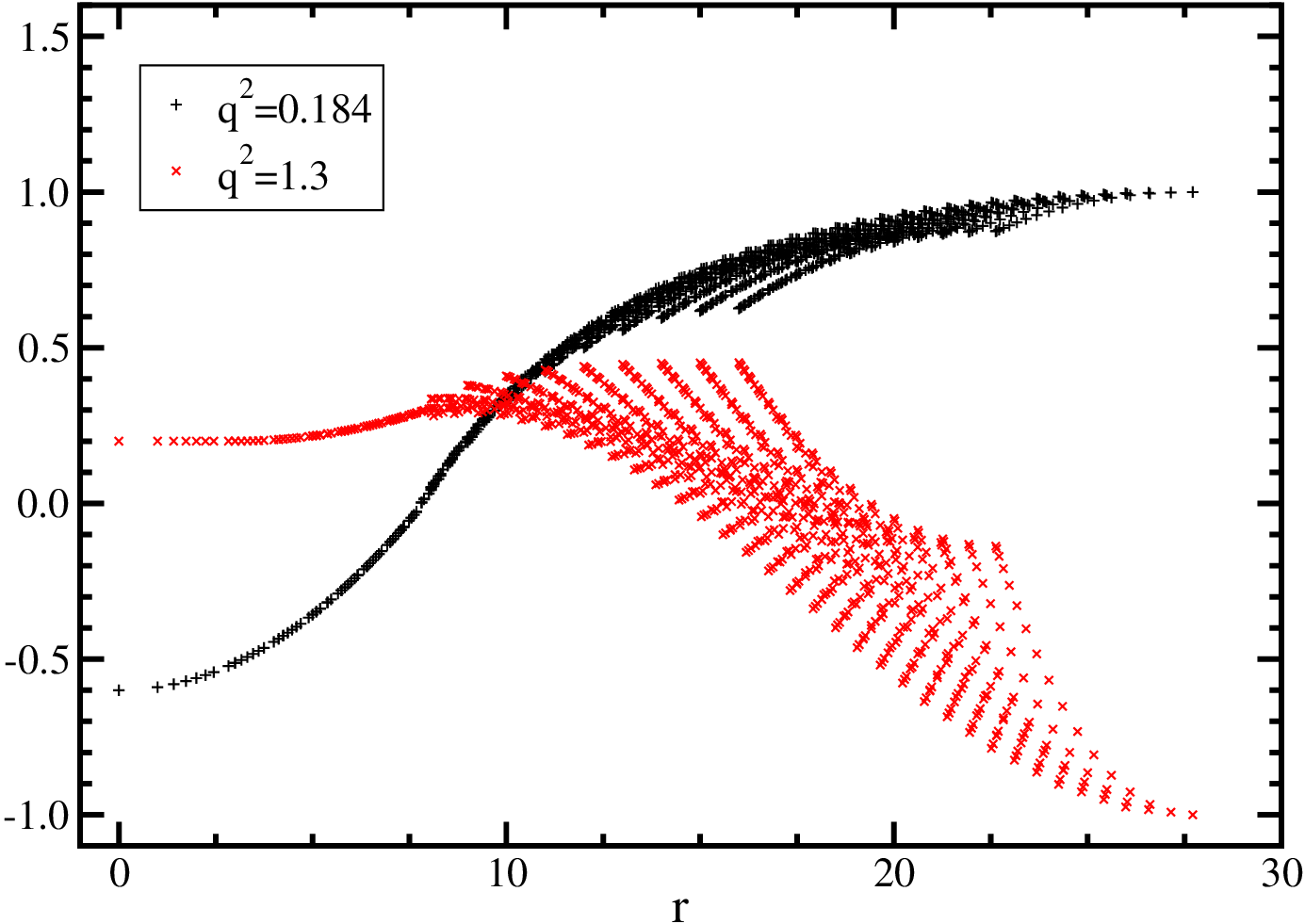}
\caption{
Smearing functions of sink operator on (4.1 fm)$^3$ box
for two-state analysis.
\label{fig:WF_32}
}
\end{figure}

\begin{figure}[!t]
\includegraphics*[angle=0,width=0.6\textwidth]{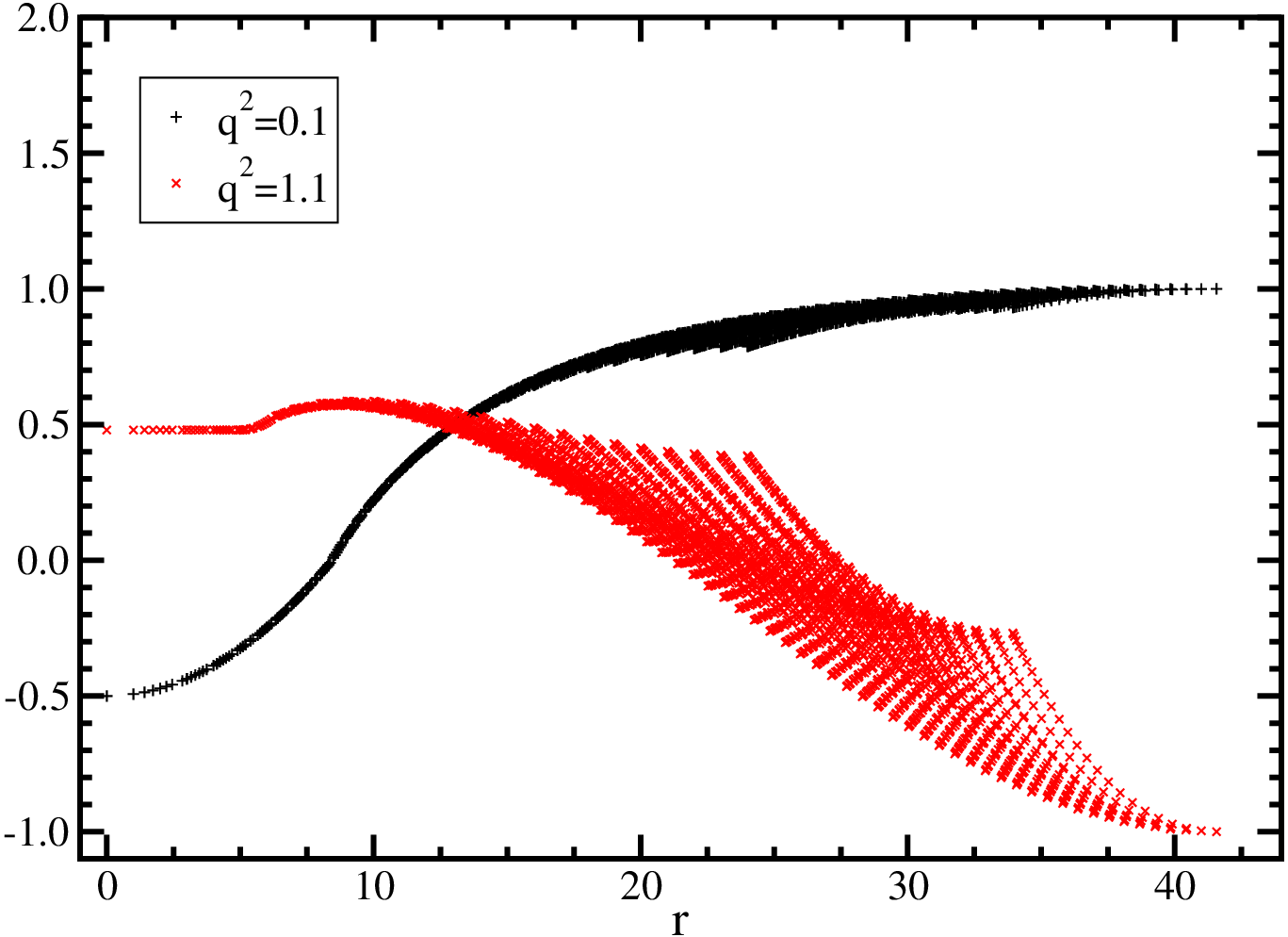}
\caption{
Same as Fig.~\ref{fig:WF_32} on (6.1 fm)$^3$ box.
\label{fig:WF_48}
}
\end{figure}

The value of $q^2$ is related to the relative momentum of 
the two-nucleon state as $p^2 = (2\pi/L)^2 \cdot q^2$, where 
$q^2$ is not an integer in general
due to the finite volume effect of 
the two-particle interaction~\cite{Luscher:1986pf,Luscher:1990ux}.
Since we need two smearing functions, 
we take  a  pair of values of $q^2$, one around zero momentum $q^2\approx 0$ and the other around unit of momentum $q^2\approx 1$ being the simplest choices,  and make trial runs to find the optimum values of $q^2$.     
Our optimization criteria are that the effective energy of one of the states is close to ground state energy obtained in the single state analysis, and that the two smearing functions have significantly different couplings 
to the ground and first excited states.   
After several trial calculations, we choose $q^2 = 0.184$ and $1.3$
for the (4.1 fm)$^3$ box, and 
$q^2 = 0.1$ and $1.1$ for the (6.1 fm)$^3$ box. 
Our  smearing functions for both volumes are plotted
in Figs.~\ref{fig:WF_32} and ~\ref{fig:WF_48}.

We note that we do not use a negative $q^2$ 
determined from the bound state
which corresponds to 
an exponentially damped smearing function. 
In the two-nucleon correlation function for the ${\cal O}_1$ source and such a sink operator,
we find  that higher excited state contributions  are not suppressed.  Hence such an operator 
is not suitable for the $2\times 2$ diagonalization of the ground and first excited states we 
attempt to carry out. 

Let us finally note that we also employ the point sink operator to carry out the single state
analysis on the second ensemble for a consistency check with the results from the first 
ensemble.

\subsection{Results for (6.1 fm)$^3$ box}

\begin{figure}[!t]
\includegraphics*[angle=0,width=0.6\textwidth]{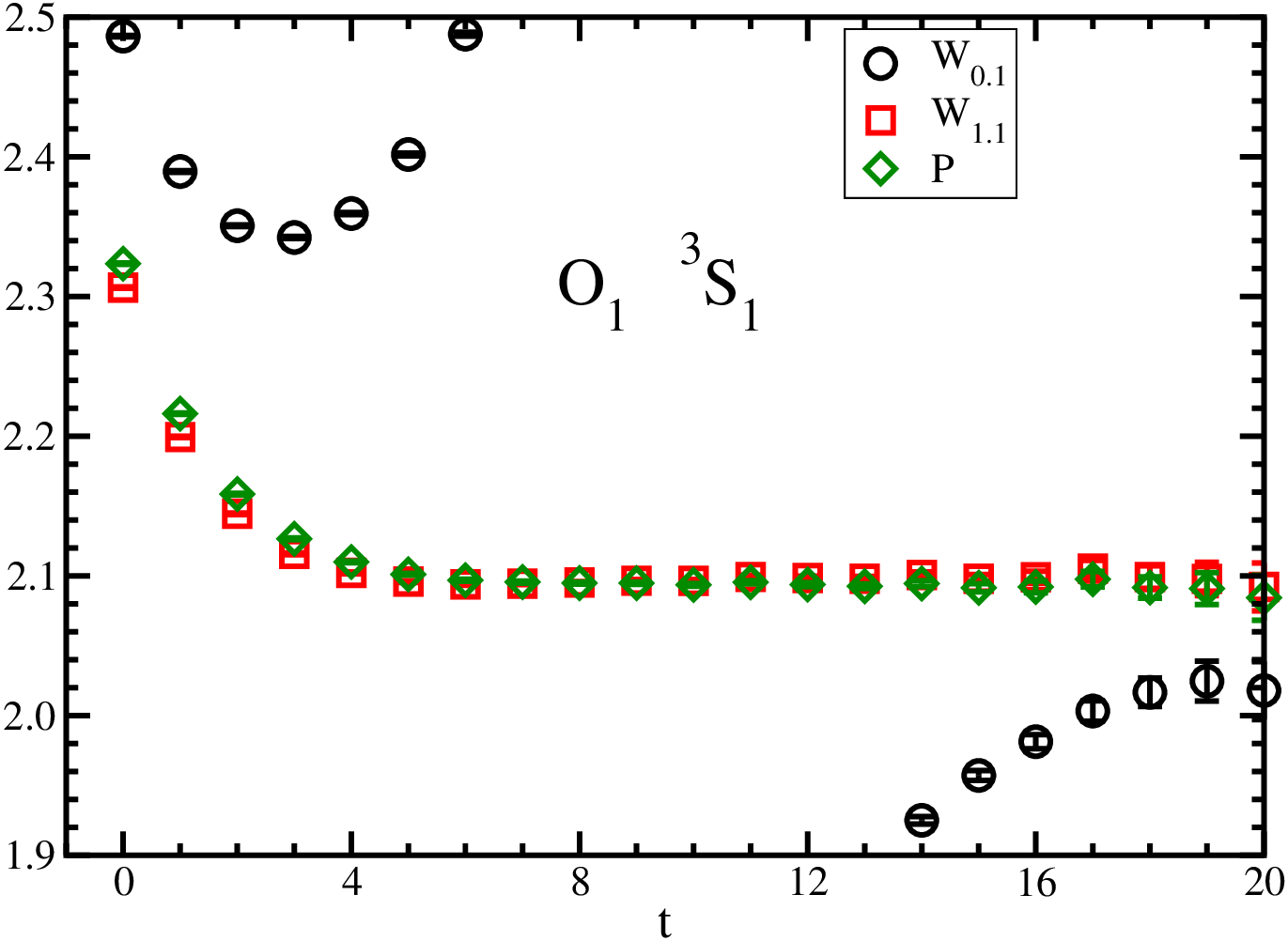}
\caption{
Effective energy of two-nucleon for $^3$S$_1$ channel 
with ${\cal O}_1$ source on (6.1 fm)$^3$ box.
Results are given in lattice units with smearing functions
$W_{0.1}$ (circle) and $W_{1.1}$ (square),
and point sink operator (diamond).
\label{fig:DN-3S1_w}
}
\end{figure}

\begin{figure}[!t]
\includegraphics*[angle=0,width=0.6\textwidth]{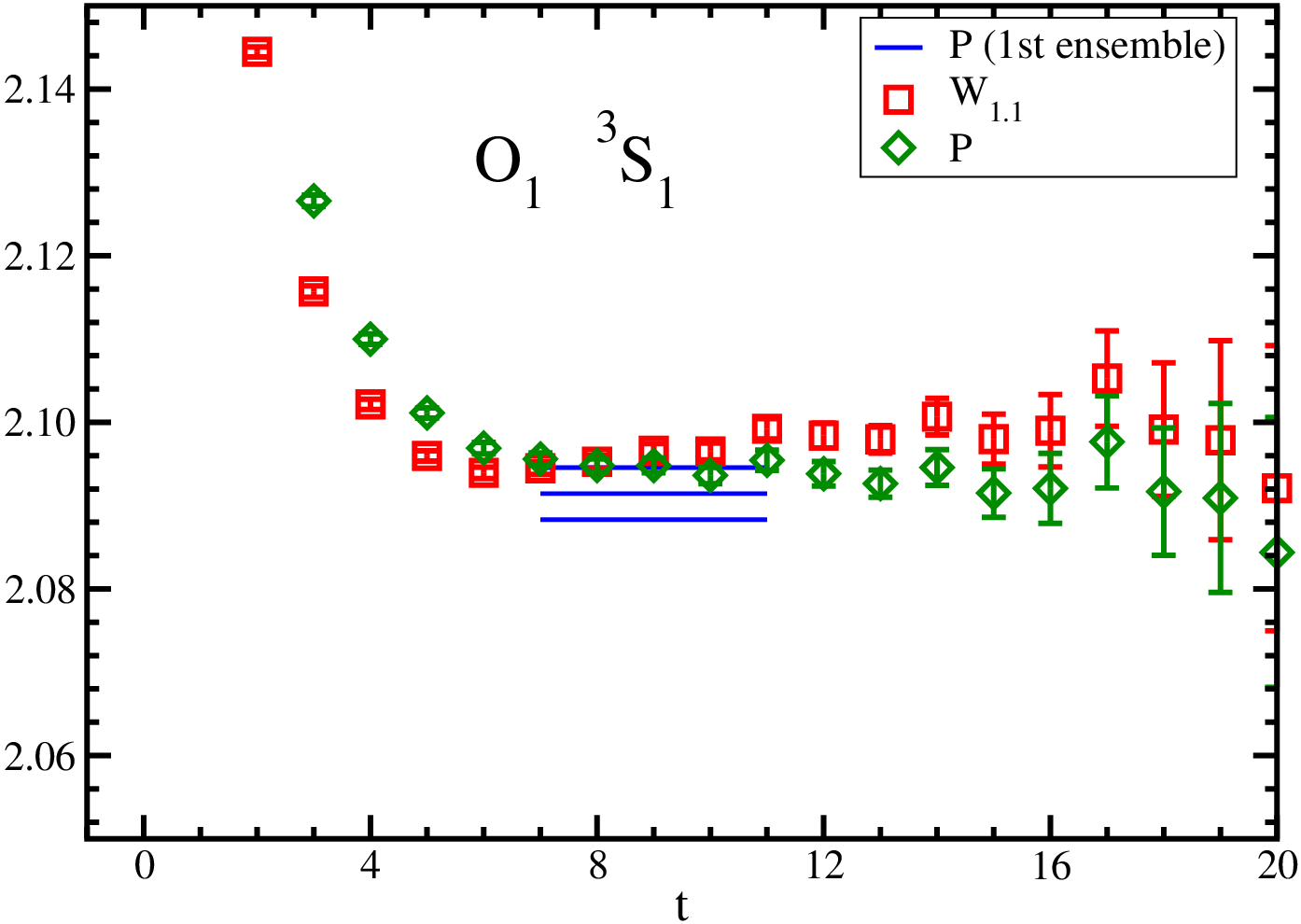}
\caption{
Same as Fig.~\ref{fig:DN-3S1_w}, but scale in vertical axis
is enlarged.
Solid lines denote fit result for ground state energy
with one standard deviation error band 
in single state analysis choosing ${\cal O}_1$ source.
\label{fig:DN-3S1}
}
\end{figure}

We first show the results for the $^3$S$_1$ channel on the (6.1 fm)$^3$ box.
Let us begin with data for the  ${\cal O}_1$ source. 
Figure~\ref{fig:DN-3S1_w} shows the effective energies in the $^3$S$_1$ 
channel for the two smearing function sinks, $W_{0.1}$ and $W_{1.1}$, and the point sink $P$. 
The effective energy for the $W_{0.1}$ smearing function
diverges around $t=8$, and rises up
from below after $t=12$ due to the sign flip of the correlation function.
Thus  at least two states
contribute to the correlation function overlapping
to the operator with different signs.
On the other hand, the result for $W_{1.1}$ is close to that for the point sink $P$.
Figure~\ref{fig:DN-3S1}  is an expanded view on this point. 

\begin{figure}[!t]
\includegraphics*[angle=0,width=0.6\textwidth]{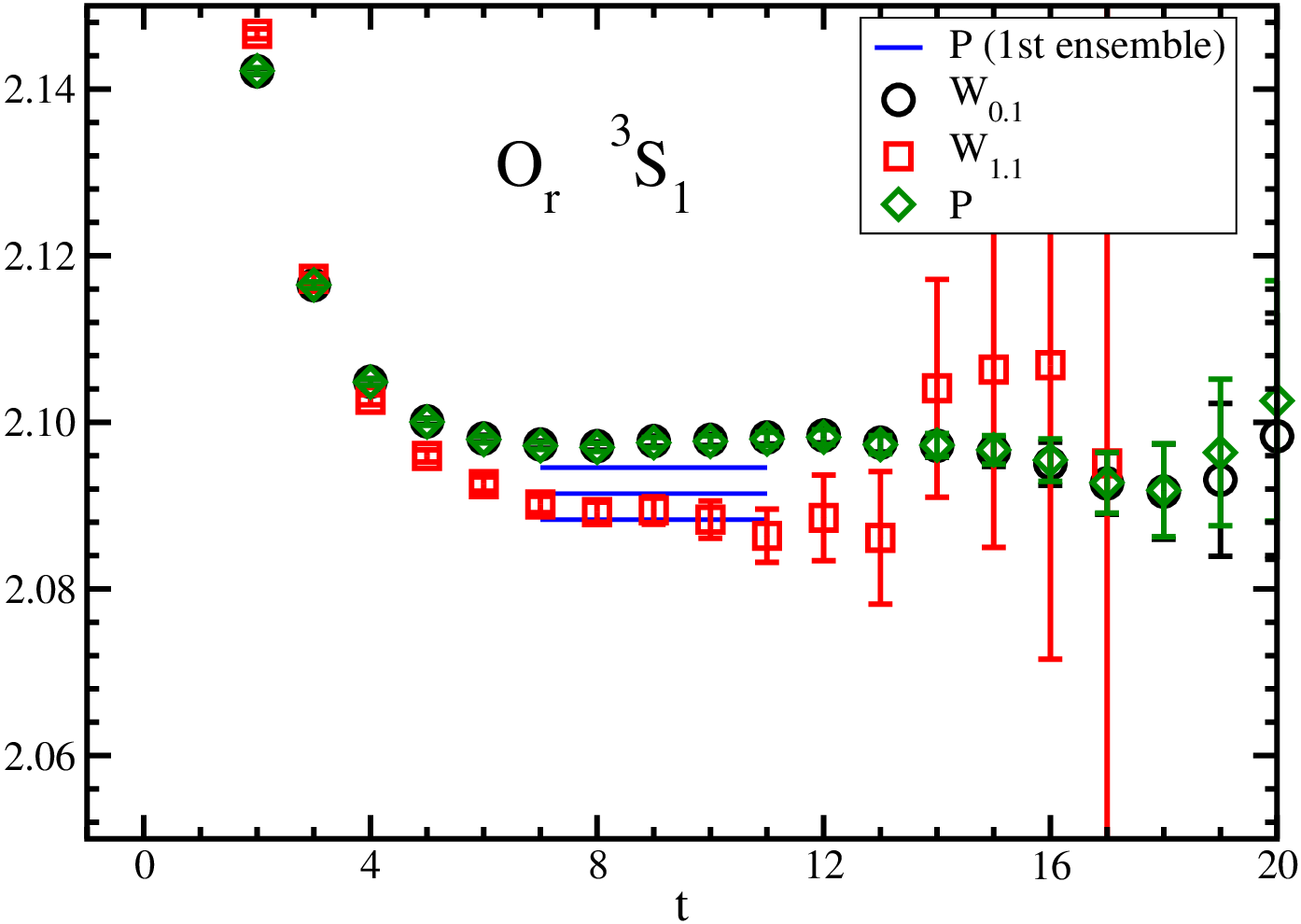}
\caption{
Same as Fig.~\ref{fig:DN-3S1} with ${\cal O}_r$ source.
\label{fig:DN2-3S1}
}
\end{figure}

Let us now look at effective energies for the ${\cal O}_r$ source in 
Fig.~\ref{fig:DN2-3S1}.  In this case the $W_{0.1}$ result is close 
to the point sink result,  whereas the $W_{1.1}$ result is lower. 

We diagonalize the following matrix at each $t$,
\begin{equation}
M(t,t_0) = C(t_0)^{-1}C(t),
\label{eq:diag_mat}
\end{equation}
where $t_0$ is a reference time and the $2\times 2$ components
of the correlation function matrix $C(t)$ are given by
\begin{equation}
C_{ij}(t) = G^{i;j}_{NN}(t)
\end{equation}
with $G^{i;j}_{NN}(t)$ being the two-nucleon correlation function
using the $i$ ($i = {\cal O}_1, {\cal O}_r$) source operator 
and the $j$ ($j = W_{0.1}, W_{1.1}$) smearing function 
for the sink operator.
With a choice of $t_0 = 6$
we determine the two eigenvalues $\lambda_\alpha(t)$ ($\alpha=0,1$) 
of $ M(t,t_0)$ at each $t$ and extract the energy of each eigenstate $\alpha$ 
through $\lambda_\alpha(t) =\exp (-\overline{E}_{L,\alpha}(t-t_0))$.

\begin{figure}[!t]
\includegraphics*[angle=0,width=0.6\textwidth]{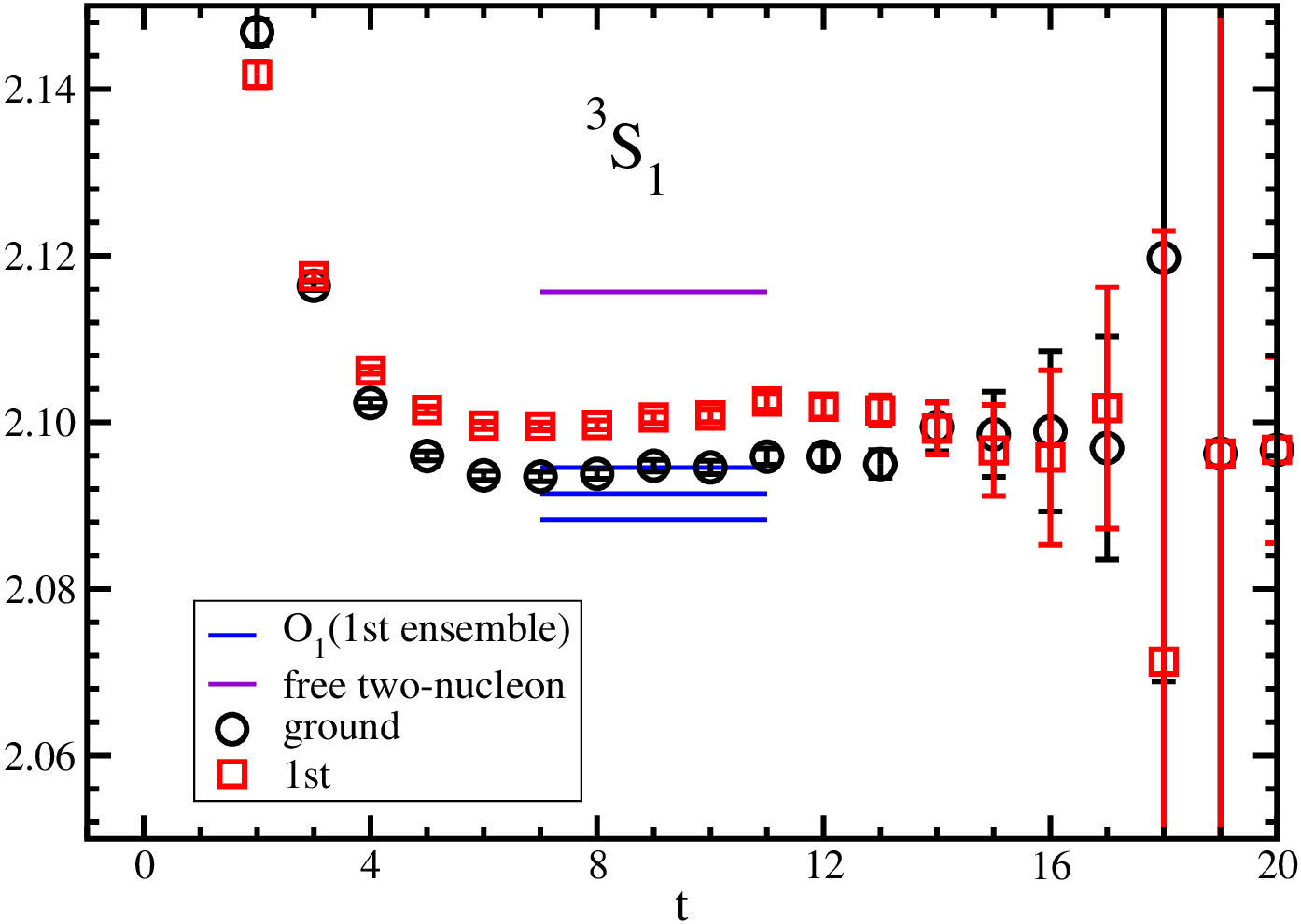}
\caption{
Effective energy of two-nucleon ground (circle) and first excited (square)
states
obtained by diagonalization method for $^3$S$_1$ channel in lattice units.
Fit result of single state analysis with ${\cal O}_1$ source
on the first ensemble 
with one standard deviation error band is expressed 
by three solid lines.
Expected energy level of free two-nucleon state with lowest relative momentum  
is denoted by single solid line.
\label{fig:ene_3S1}
}
\end{figure}

The effective energies of the eigenstates obtained from the diagonalization
are plotted in Fig.~\ref{fig:ene_3S1}.
The energies for the two states are clearly separated in the plateau region.
The ground state result is reasonably consistent with the result of 
the single state analysis with the ${\cal O}_1$ source 
obtained on the first ensemble, which is expressed by the three solid lines in
the figure.
The first excited state is clearly higher than the ground state, but it 
is much lower than the free case with the lowest relative momentum, whose
energy is given by $2\sqrt{m_N^2 + (2\pi/L)^2}$ 
denoted by the single solid line in the figure.

In order to determine the energy shift as in Sec.~\ref{sec:1st},
we define the ratio of the eigenvalue obtained from the diagonalization
to the nucleon correlation function squared,
\begin{equation}
\overline{R}_{\alpha}(t) = 
\frac{\lambda_{\alpha}(t)}{\left(G_N(t)\right)^2}.
\label{eq:R-bar}
\end{equation}
We also define the effective energy shift of the ratio 
$\overline{R}_{\alpha}$ as,
\begin{equation}
\Delta \overline{E}_{L,\alpha}^{\mathrm{eff}} = 
\ln \left(\frac{\overline{R}_{\alpha}(t)}
{\overline{R}_{\alpha}(t+1)}\right).
\end{equation}

\subsubsection{Ground state in the $^3$S$_1$ channel} 

\begin{figure}[!t]
\includegraphics*[angle=0,width=0.6\textwidth]{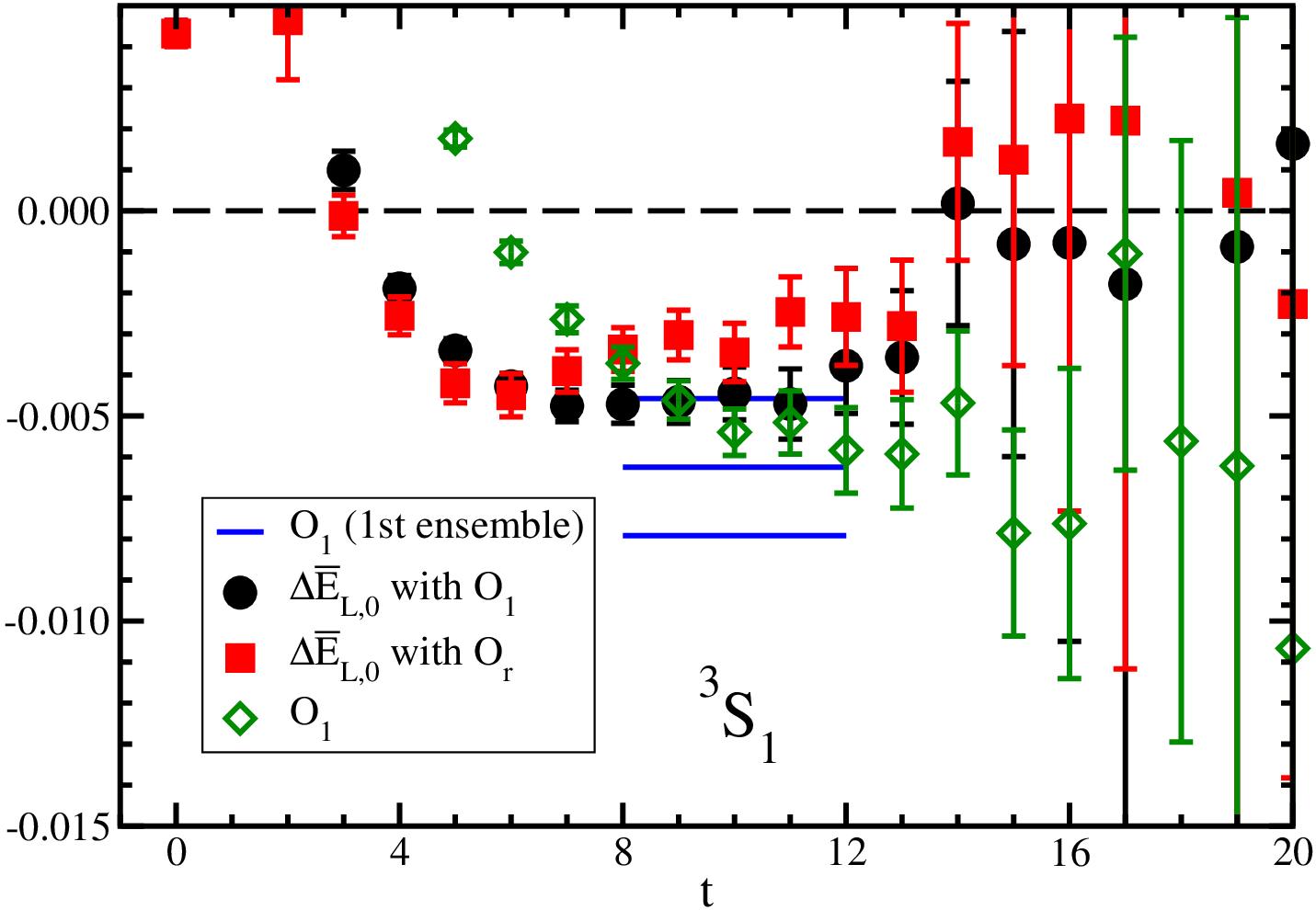}
\caption{
Effective energy shifts of two-nucleon  
$\Delta \overline{E}_{L,0}^{\mathrm{eff}}$ for $^3$S$_1$ channel 
on (6.1 fm)$^3$ box in lattice units.
Nucleon correlators with
${\cal O}_1$ (circle) and ${\cal O}_r$ (square) sources are employed
in the denominator of Eq.~(\ref{eq:R-bar}).
Result of single state analysis 
with ${\cal O}_1$ source $\Delta E_{L}^{\mathrm{eff}}$
is also plotted by open diamond symbol.
Fit result of single state analysis with ${\cal O}_1$ source
on the first ensemble
with one standard deviation error band is given 
by solid lines.
\label{fig:eff_R_3S1_n0}
}
\end{figure}

Figure~\ref{fig:eff_R_3S1_n0} shows a compilation of all data for the ground state 
both from the diagonalization analysis as well as from the single state analysis. 
The solid circles represent the effective energy shift of the ground state 
$\Delta \overline{E}_{L,0}^{\mathrm{eff}}$ using the ${\cal O}_1$ source 
in the nucleon propagator in the denominator of Eq.~(\ref{eq:R-bar}).  
The solid squares are the ones using the ${\cal O}_r$  source in the nucleon 
propagator.  The diamonds show the energy shift from single state analysis 
using point sink, but based on the second ensemble.  Finally the three lines 
show the estimated ground state energy shift from the single state analysis 
of the ${\cal O}_1$ source with 
the point sink from the first ensemble. 

\begin{figure}[!t]
\includegraphics*[angle=0,width=0.6\textwidth]{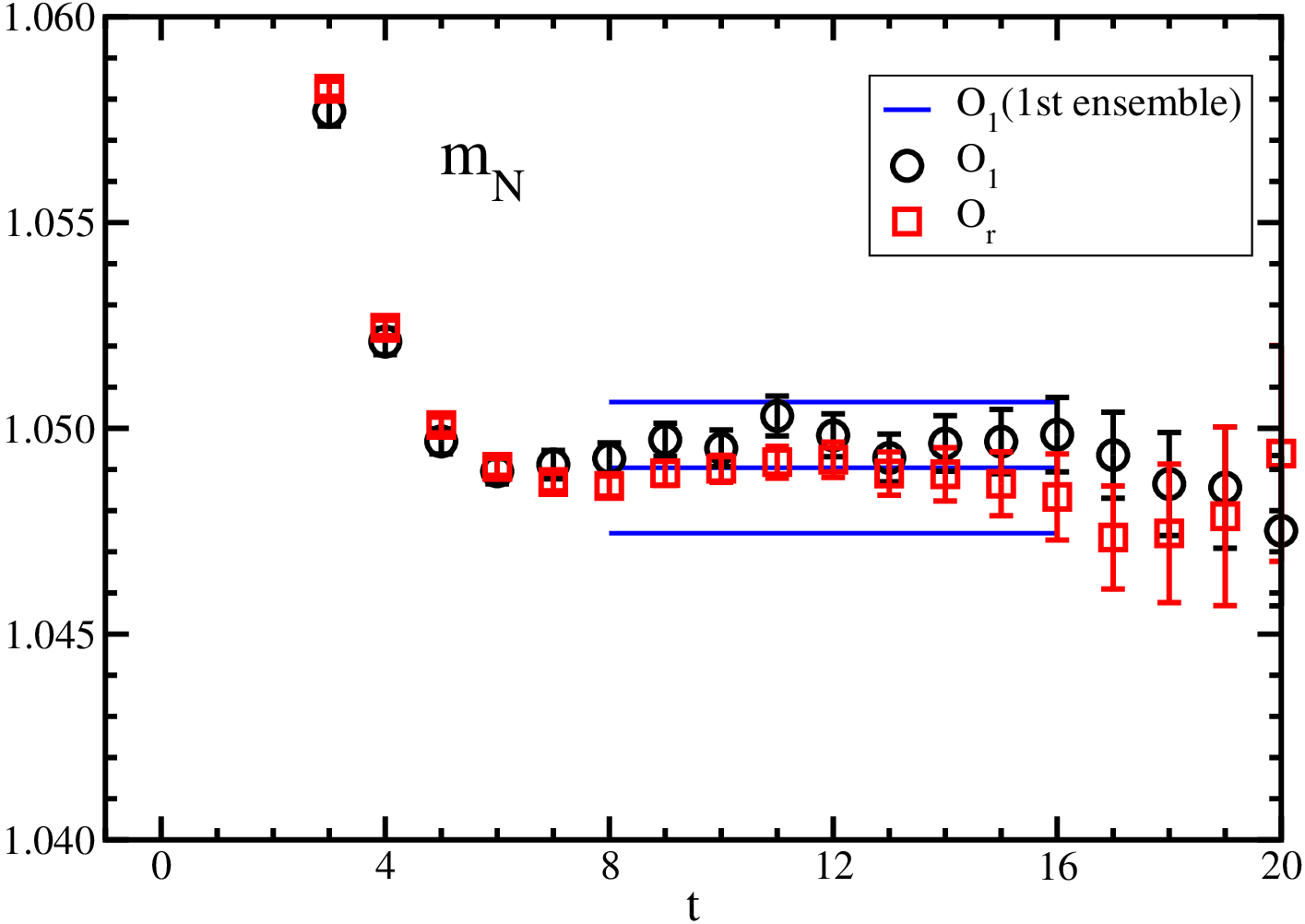}
\caption{
Nucleon effective mass on (6.1 fm)$^3$ box for the second ensemble with
${\cal O}_1$ (circle) and ${\cal O}_r$ (square) sources in lattice units.
Fit result of ${\cal O}_1$ source on the first ensemble 
with one standard deviation error band is expressed 
by solid lines.
\label{fig:eff_N}
}
\end{figure}

We find it gratifying that the diagonalization results for the ground state exhibit
clear plateaux over a significant time range extending from $t=7$.   
A somewhat higher value of the 
plateau if one takes the  ${\cal O}_r$ source in the nucleon propagator can be traced back 
to a systematic shift in the nucleon effective mass itself, see Fig.~\ref{fig:eff_N}, 
so that the difference should be regarded as a measure of systematic error. 
The plateaux are also consistent with the result of the single state analysis from 
the same ensemble (diamonds), which in turn 
are also consistent with that from the first ensemble (solid lines). 

We determine the central value of the energy shift from the exponential
fit of the $\overline{R}_{0}(t)$ using the nucleon correlation
function of the ${\cal O}_1$ source with the fit range of $t=7$--13.
The systematic error due to estimate of the threshold $2m_N$ is made from the difference between
the two results with the ${\cal O}_1$ and ${\cal O}_r$ sources
for the nucleon correlators
in the denominator of the $\overline{R}_{0}$.
The systematic error associated with 
the fit range is estimated by changing the maximum
or minimum time slice of the fit range by $\pm 1$.
Table~\ref{tab:DdE} summarizes the numerical values
for the energy shift from the diagonalization analysis $\Delta \overline{E}_{L,0}$, 
and the single state analysis $\Delta E_L$.  The statistical
and systematic errors are presented in the first and second parentheses,
respectively.
We employ an  asymmetric systematic error for 
$\Delta \overline{E}_{L,0}$ to properly reflect an upward shift for the ${\cal O}_r$ source 
relative to the  ${\cal O}_1$ source.

\subsubsection{First excited state in the $^3$S$_1$ channel}

\begin{figure}[!t]
\includegraphics*[angle=0,width=0.6\textwidth]{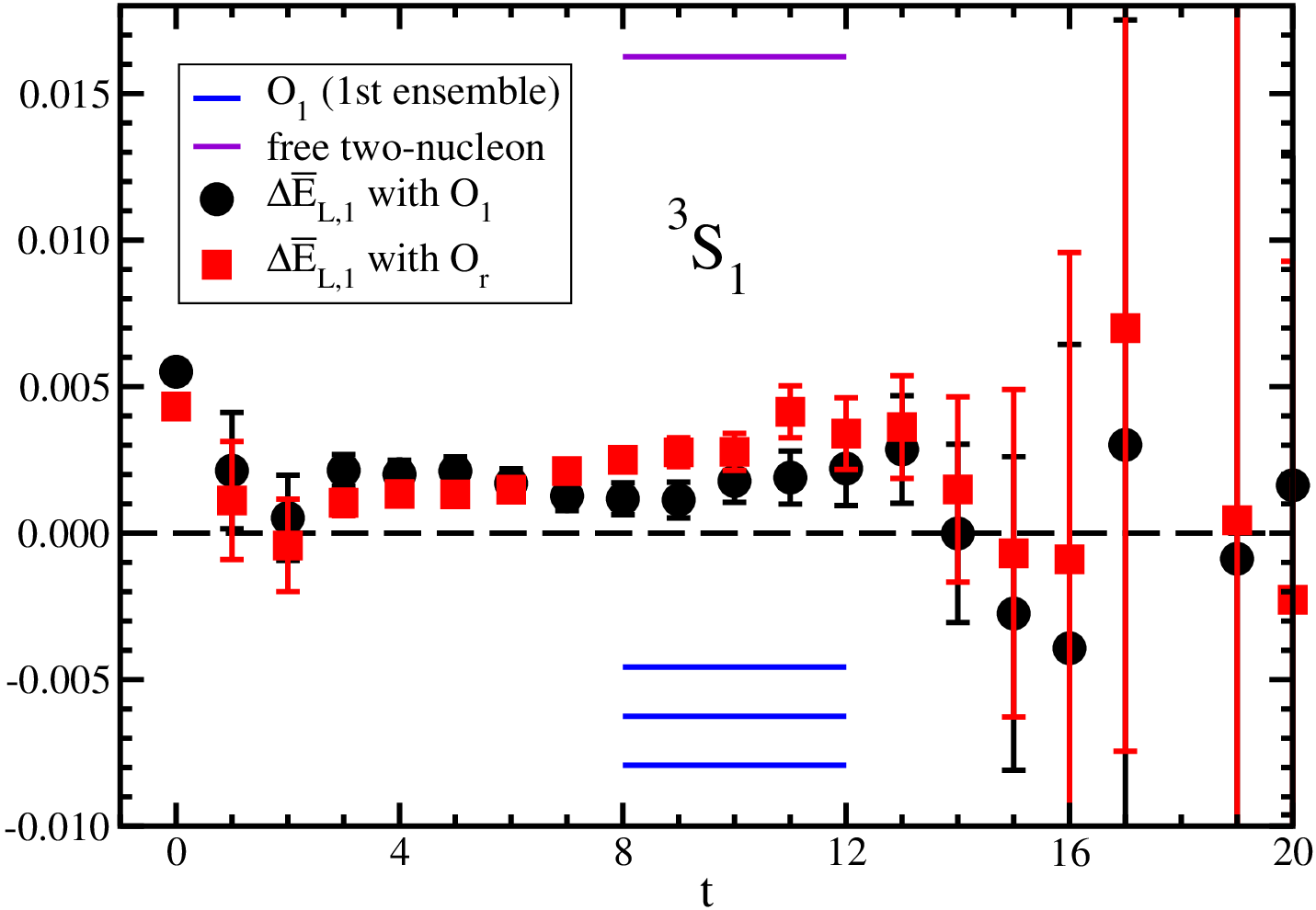}
\caption{
Same as Fig.~\ref{fig:eff_R_3S1_n0} for the first excited state.
\label{fig:eff_R_3S1_n1}
Expected energy level of free two-nucleon state 
with lowest relative momentum 
is denoted by single solid line.
}
\end{figure}

Figure~\ref{fig:eff_R_3S1_n1} shows 
the effective energy shift of the first excited state 
$\Delta \overline{E}_{L,1}^{\mathrm{eff}}$.  
Once again, we find a long plateau for both  ${\cal O}_1$ and  ${\cal O}_r$ whose 
values are mutually consistent.  
The very important feature is that the plateaux are definitively above the threshold 
and significantly lower than the value expected from the free two-nucleon state with unit relative 
momentum as presented by the single solid line in the figure.
This is consistent with the ground state being a bound state.  

The energy shift $\Delta \overline{E}_{L,1}$ is determined
from an exponential fit of the $\overline{R}_{1}(t)$
using the ${\cal O}_1$ source nucleon correlator 
with the fit range of $t=7$--13.
We use the fit result for the ${\cal O}_r$ source nucleon correlator 
to estimate a systematic error of the energy shift.
The numerical value is given in Table~\ref{tab:DdE_n1} .

\subsubsection{Analysis of the  $^1$S$_0$ channel}

\begin{figure}[!t]
\includegraphics*[angle=0,width=0.6\textwidth]{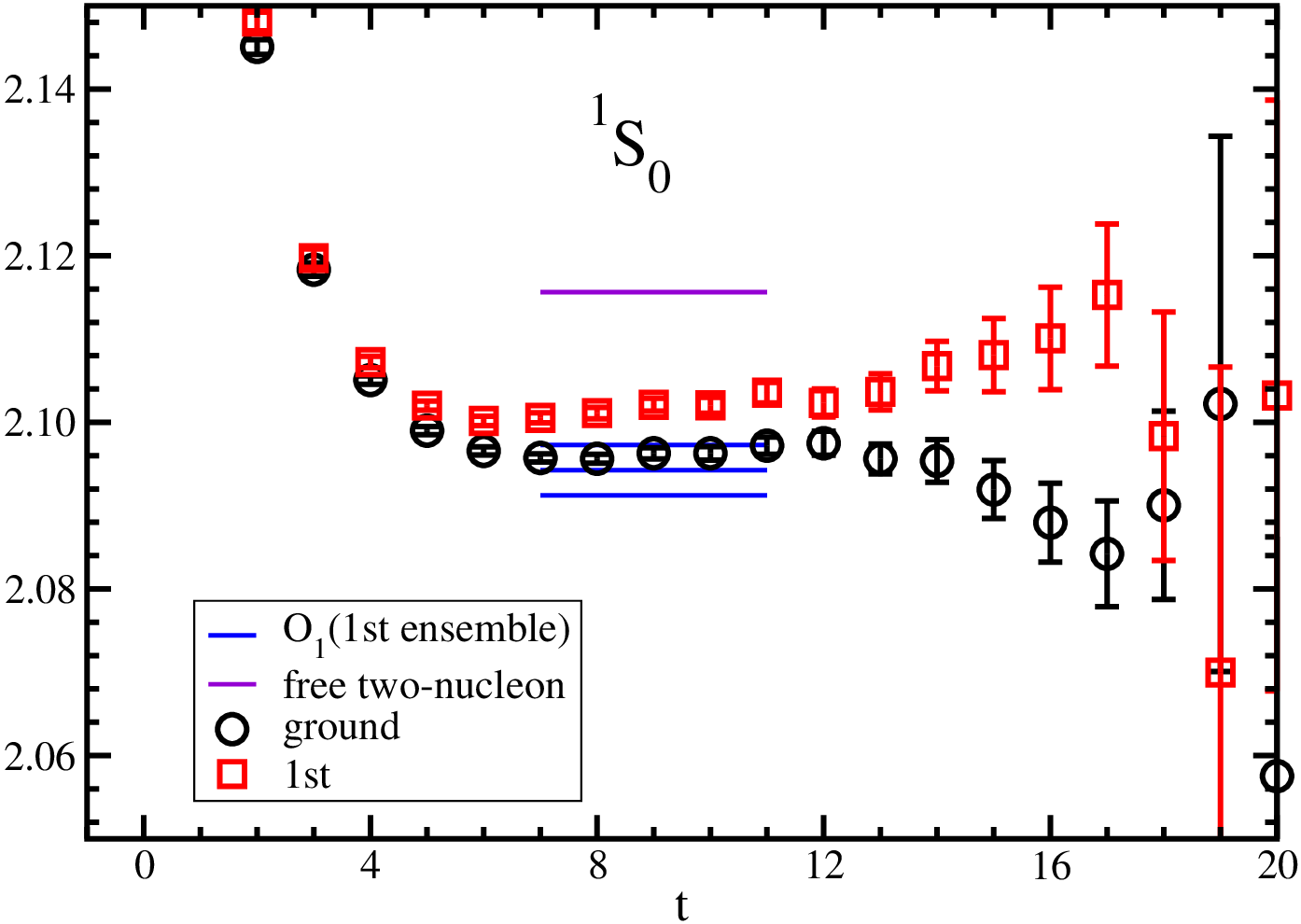}
\caption{
Same as Fig.~\ref{fig:ene_3S1} for $^1$S$_0$ channel.
\label{fig:ene_1S0}
}
\end{figure}

In the $^1$S$_0$ channel, the behaviors of the two-nucleon 
correlation functions are similar to those in the $^3$S$_1$ channel,
so that we will present only the results after the diagonalization.
The effective energies of the eigenstates are shown in Fig.~\ref{fig:ene_1S0}.
The signals are clean and both results show clear plateaux.
We observe that the ground state energy is consistent with the
result on the first ensemble of (6.1 fm)$^3$ box 
denoted by the three solid lines.

\begin{figure}[!t]
\includegraphics*[angle=0,width=0.6\textwidth]{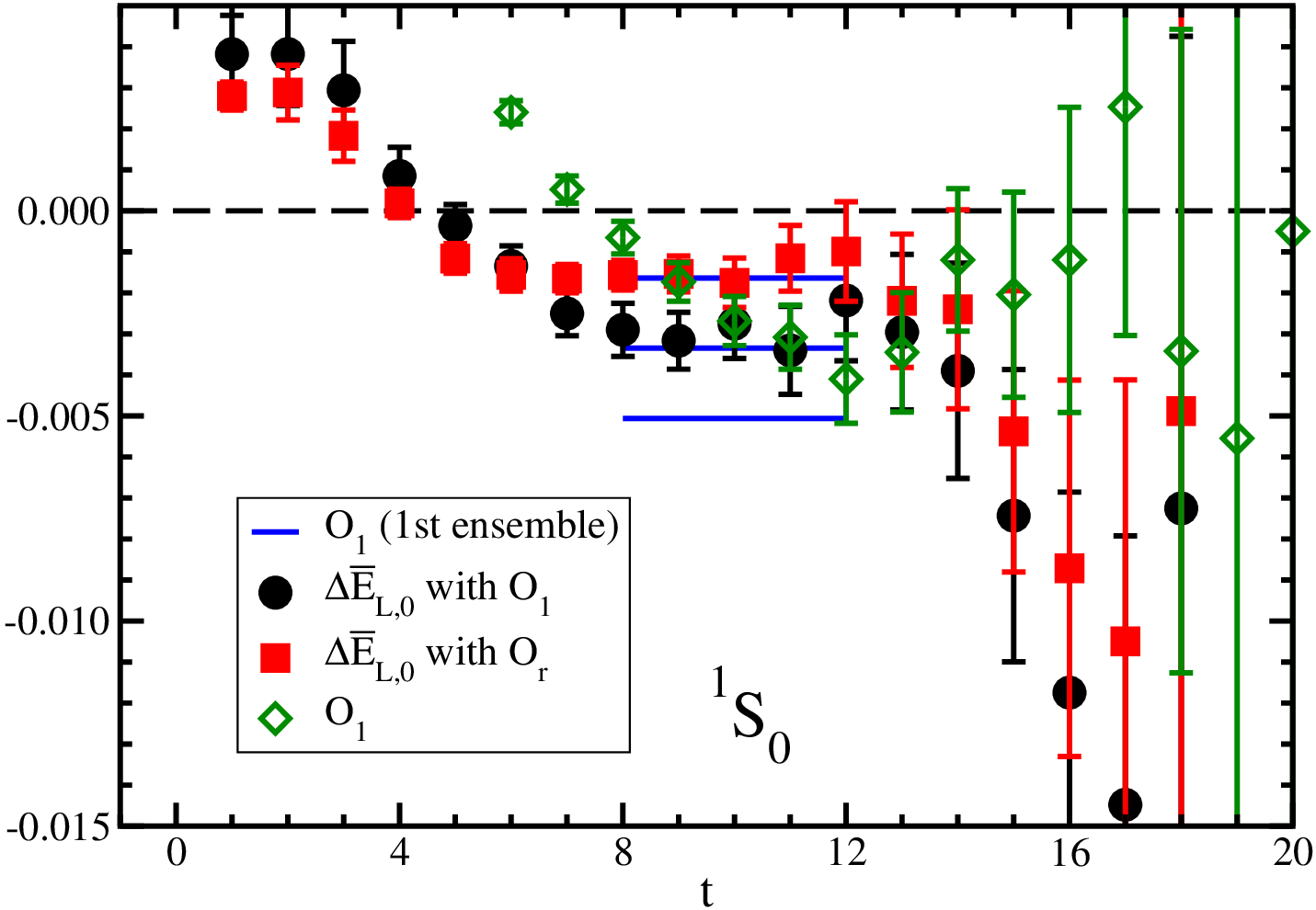}
\caption{
Same as Fig.~\ref{fig:eff_R_3S1_n0} for $^1$S$_0$ channel.
\label{fig:eff_R_1S0_n0}
}
\end{figure}

\begin{figure}[!t]
\includegraphics*[angle=0,width=0.6\textwidth]{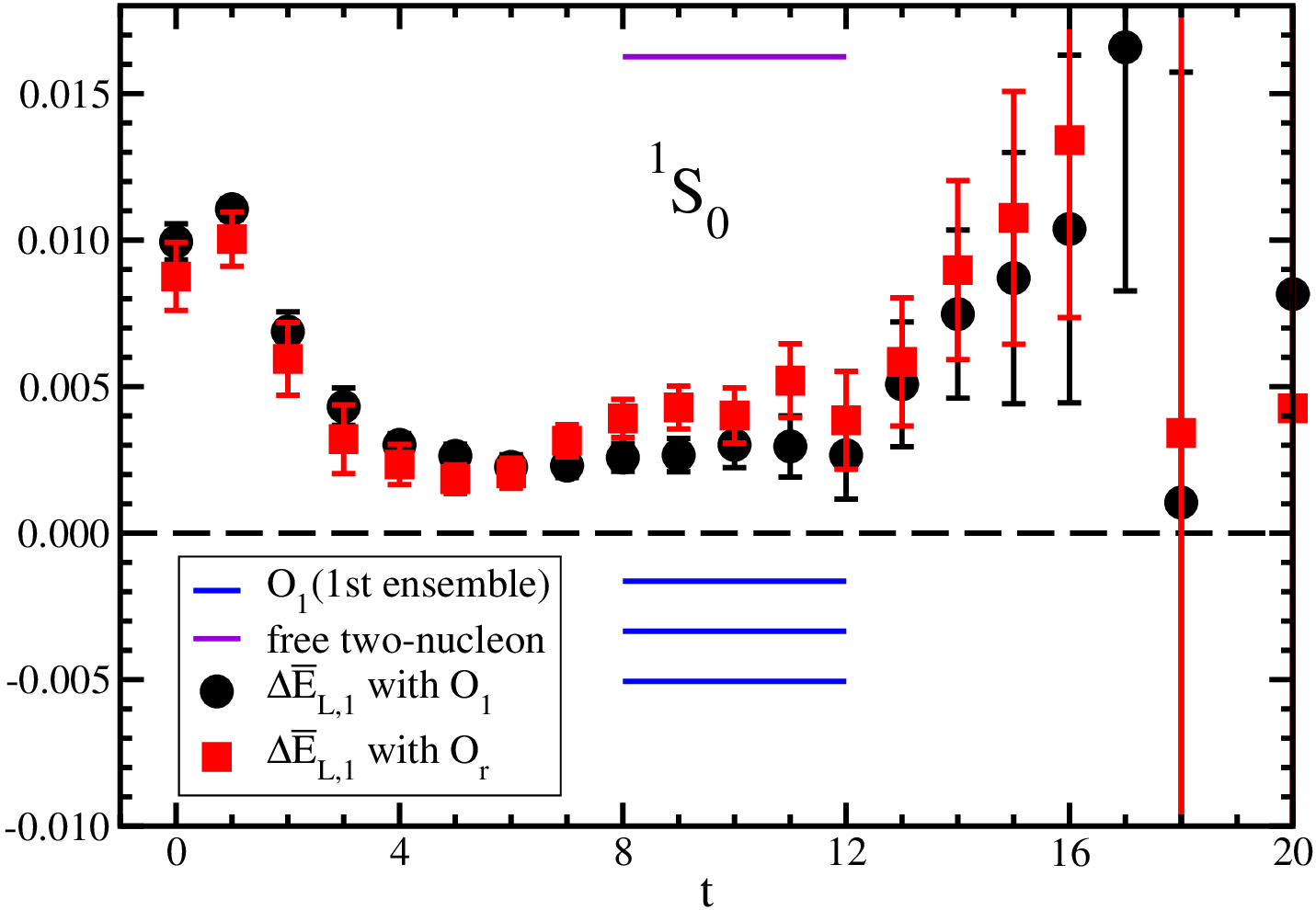}
\caption{
Same as Fig.~\ref{fig:eff_R_3S1_n1} for $^1$S$_0$ channel.
\label{fig:eff_R_1S0_n1}
}
\end{figure}

In Figs.~\ref{fig:eff_R_1S0_n0} and \ref{fig:eff_R_1S0_n1}
the effective energy shift for the ground and first excited states
$\Delta \overline{E}_{L,\alpha}^{\mathrm{eff}}$ are respectively shown as well as the result
of $\Delta E_{L}^{\mathrm{eff}}$ for the ground state calculated on the second ensemble.
We find features similar to those in the $^3$S$_1$ channel,  
including long plateaux and  systematic biases due to the choice of the 
source operators.  The results for energy shift for the ground and first excited states
are summarized in Tables~\ref{tab:DdE} and \ref{tab:DdE_n1}, respectively, 
where the errors are estimated as in the ${}^3S_1$ channel.

We observe that the absolute value of the energy shift of the ground state
is almost half of that in the 
$^3$S$_1$ channel.  This is consistent with the observation in the first
calculation.
On the other hand,  the energy shift of the first excited state  
shown in Fig.~\ref{fig:eff_R_1S0_n1} is 
almost twice larger than that in the $^3$S$_1$ channel in
Fig.~\ref{fig:eff_R_3S1_n1}.
This finding is consistent with the property of a system which contains
a shallow bound state: The scattering length negatively increases as the binding energy decreases,  diverging when the binding energy vanishes.

We confirm then that the two-nucleon system in the 
$^1$S$_0$ channel at the heavy quark mass of $m_\pi = 0.8$ GeV
has a bound state as in the $^3$S$_1$ channel.

\subsection{Results for (4.1 fm)$^3$ box}
 
Scattering states have sensitive dependence on the spatial volume whereas
bound states do not change much once the spatial size is sufficiently large to 
contain them.  We repeated the diagonalization analysis on a (4.1 fm)$^3$ box
to examine if such a difference of the two types of states can be confirmed for the ground 
and first excited states in our case.

\begin{figure}[!t]
\includegraphics*[angle=0,width=0.6\textwidth]{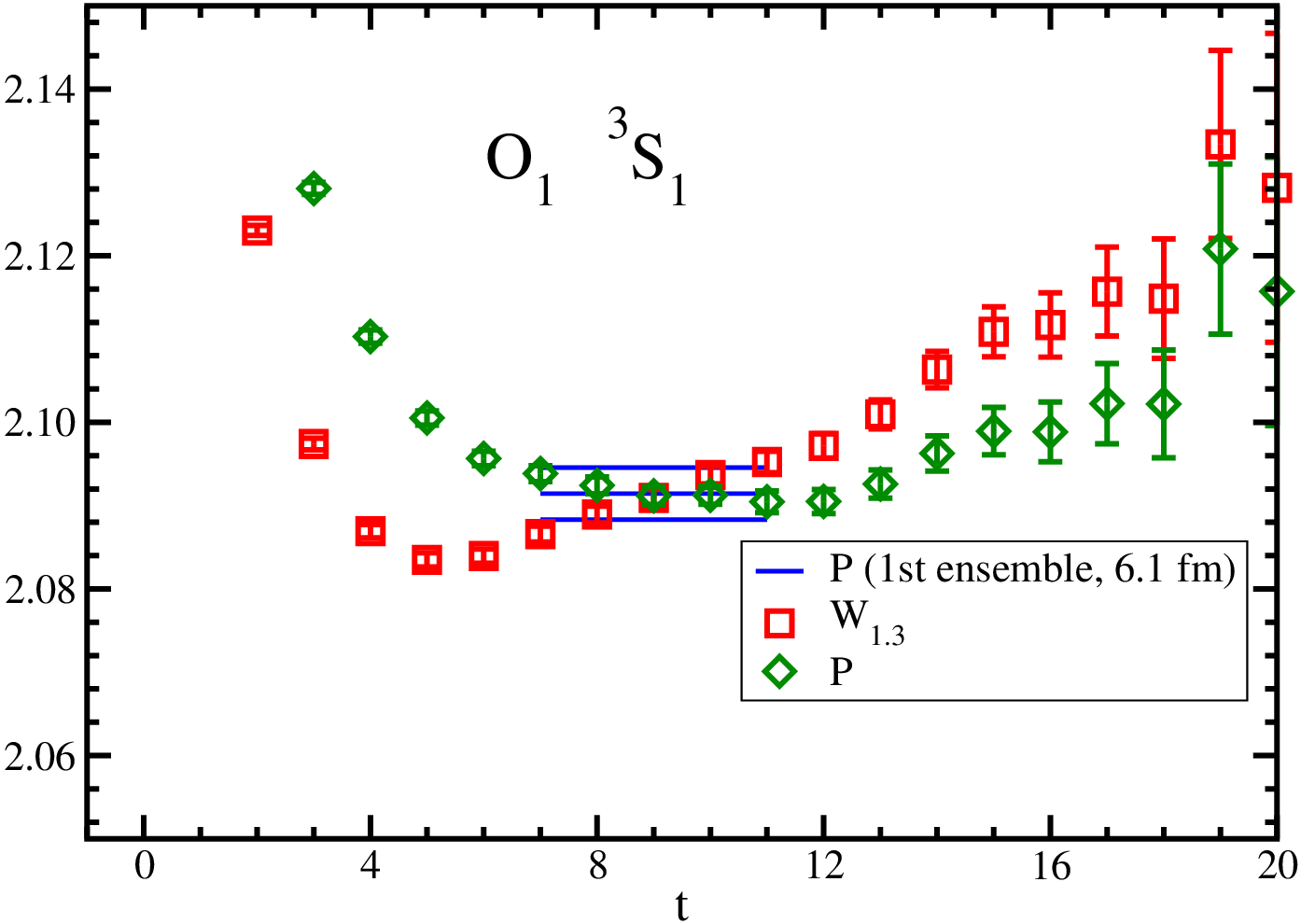}
\caption{
Effective energy of two-nucleon of ${\cal O}_1$ source 
on (4.1 fm)$^3$ box for
$^3$S$_1$ channel in lattice units with smearing function $W_{1.3}$ (square)
and point sink operator (diamond).
Fit result of single state analysis with ${\cal O}_1$ source
on the first ensemble of (6.1 fm)$^3$
with one standard deviation error band is denoted 
by solid lines.
\label{fig:DN-3S1_32}
}
\end{figure}

\begin{figure}[!t]
\includegraphics*[angle=0,width=0.6\textwidth]{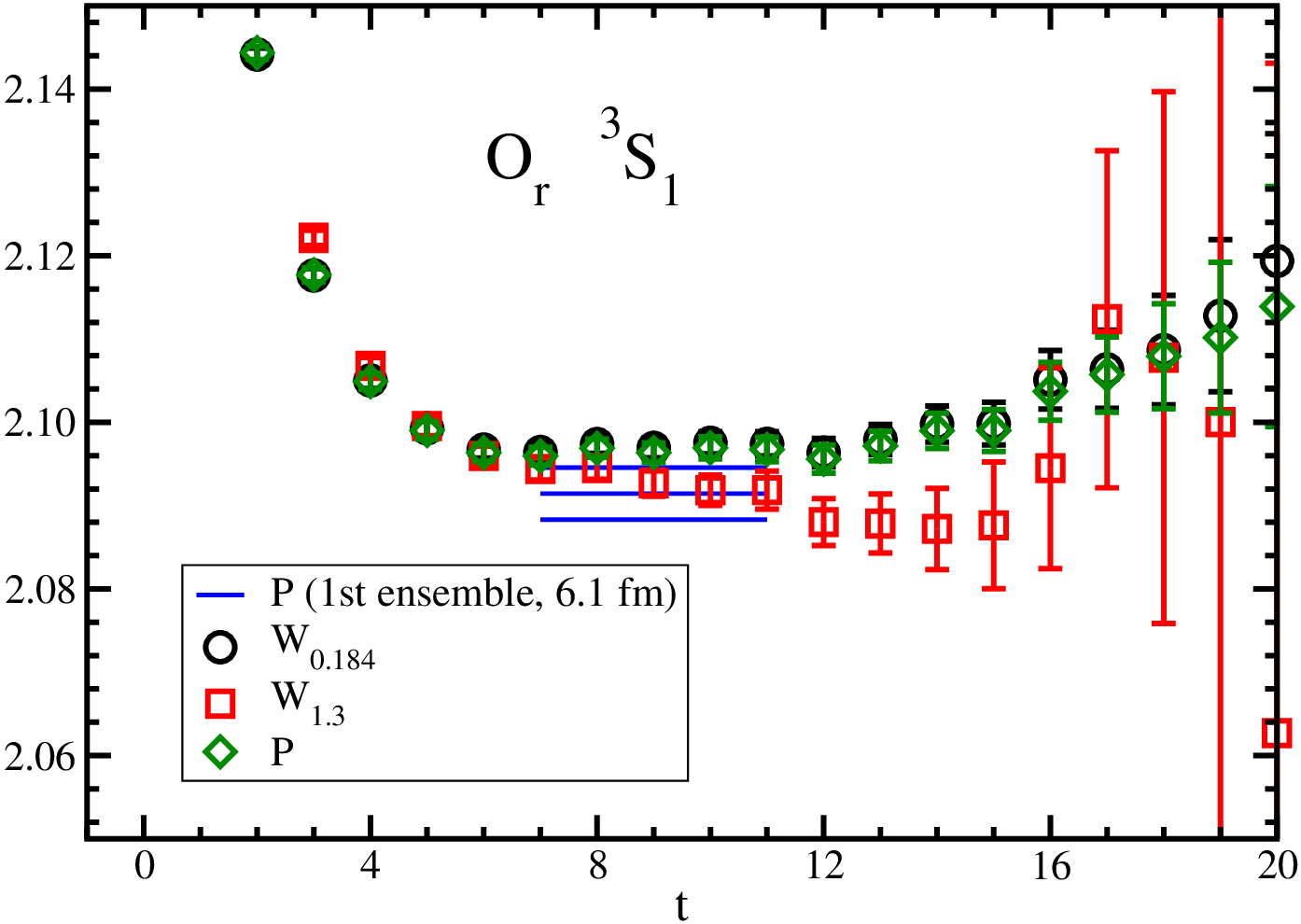}
\caption{
Same as Fig.~\ref{fig:DN-3S1_32} with ${\cal O}_r$ source.
\label{fig:DN2-3S1_32}
}
\end{figure}

The effective energies of the two-nucleon correlation functions 
with the ${\cal O}_1$ and ${\cal O}_r$ source operators for the $^3$S$_1$ channel
are plotted in Figs.~\ref{fig:DN-3S1_32} and ~\ref{fig:DN2-3S1_32}, 
respectively.
The behavior we observe is similar to 
the case of the (6.1 fm)$^3$ box except that 
the effective energy with the smearing function $W_{1.3}$ has a visible
slope in the region where the point sink result shows a plateau. 

\begin{figure}[!t]
\includegraphics*[angle=0,width=0.6\textwidth]{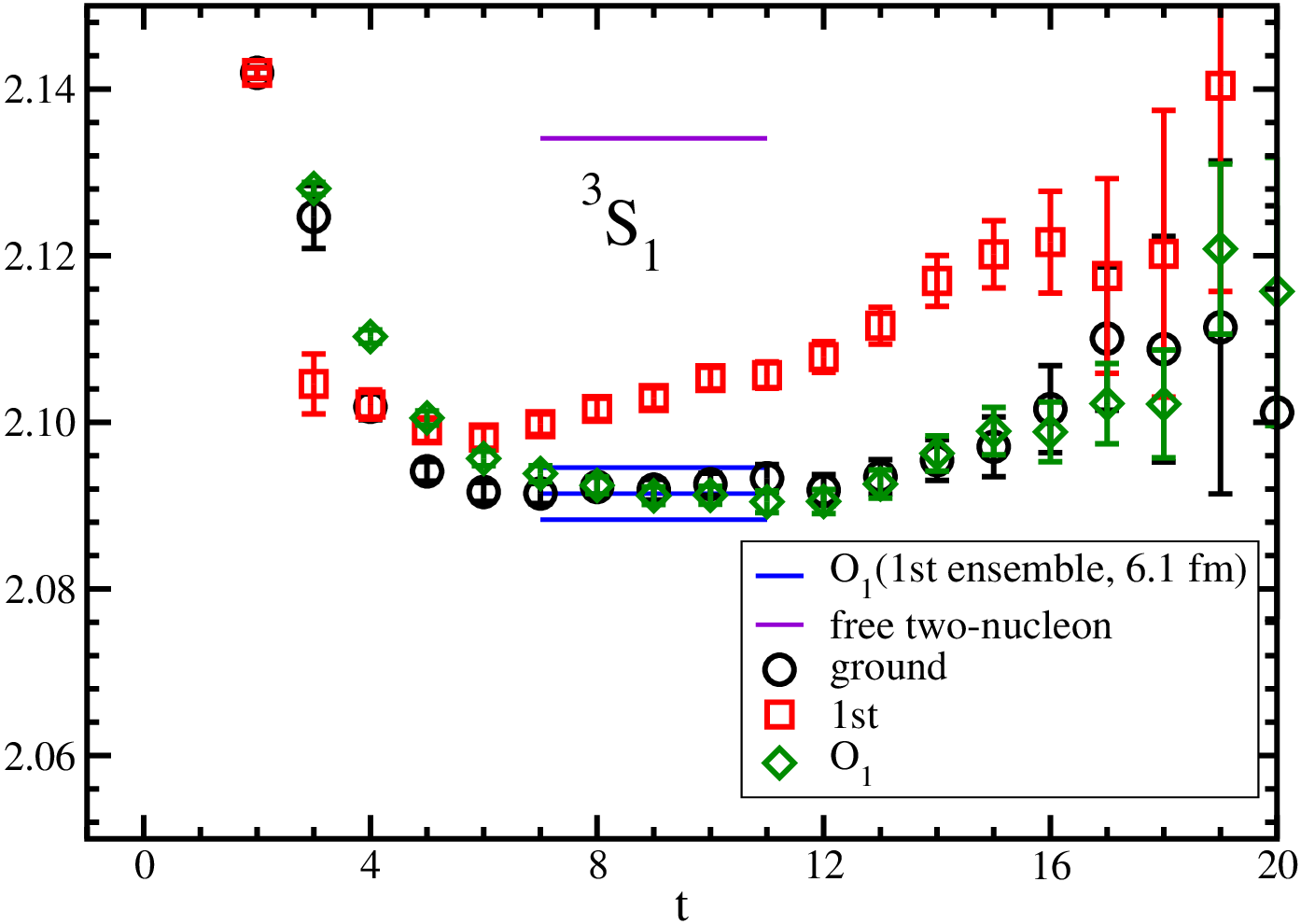}
\caption{
Same as Fig.~\ref{fig:ene_3S1} on (4.1 fm)$^3$ box.
Result of single state analysis is denoted by diamond symbol.
\label{fig:ene_3S1_32}
}
\end{figure}

Figure~\ref{fig:ene_3S1_32} presents the diagonalization results for the 
two-nucleon effective energy employing $t_0 = 8$ for the reference time 
in Eq.~(\ref{eq:diag_mat}).  For the ground state, it is once again gratifying to 
find a plateau over a sizable range of time, with the value consistent with 
that from the single state analysis.  However, the effective energy for the first 
excited state exhibits a visible slope, which was not seen in the (6.1 fm)$^3$ box 
case. 
The positive slope in the first excited state indicates the
presence of contaminations from higher excited state in the correlation
functions.

\begin{figure}[!t]
\includegraphics*[angle=0,width=0.6\textwidth]{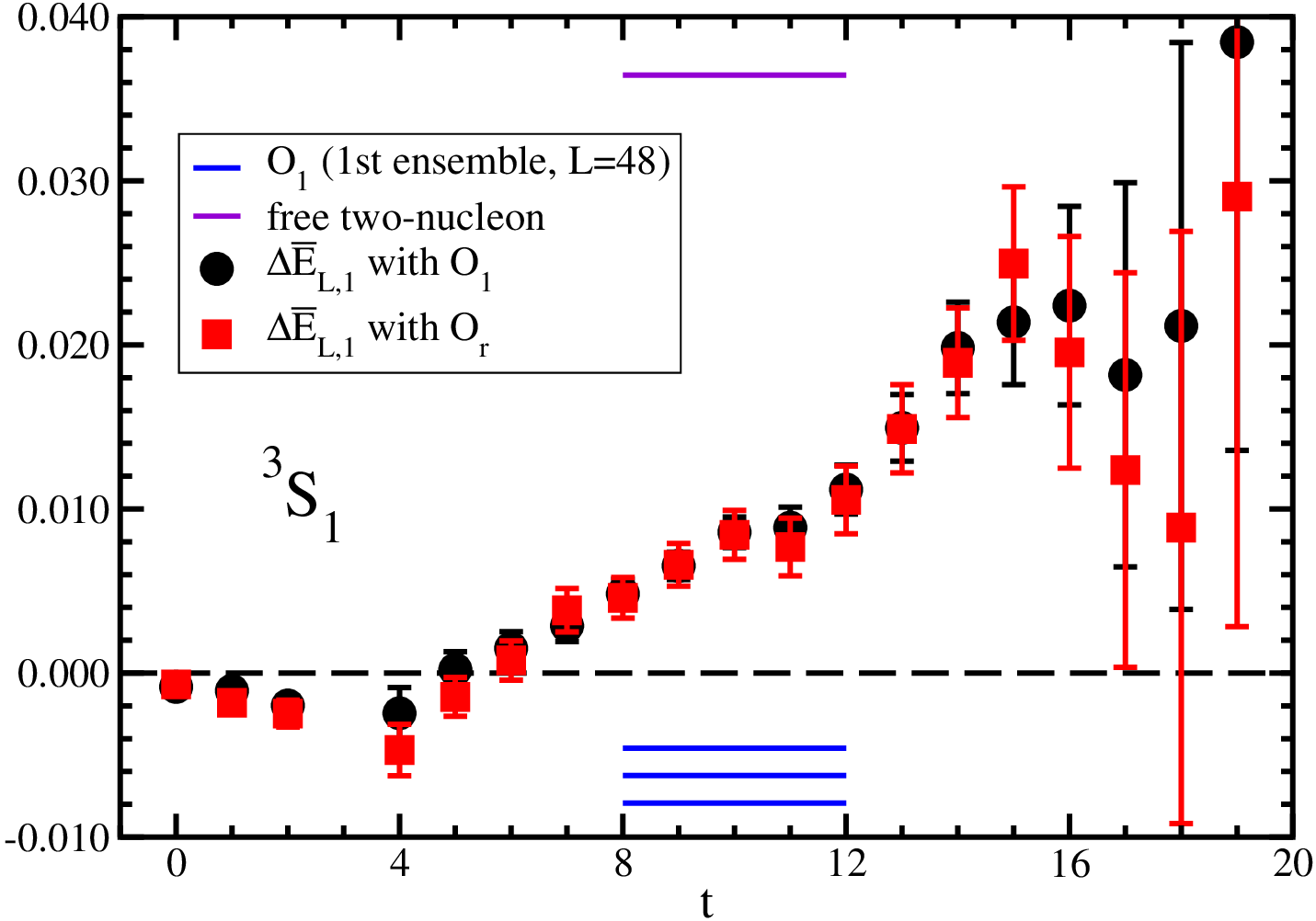}
\caption{
Same as Fig.~\ref{fig:eff_R_3S1_n1} on (4.1 fm)$^3$ box.
\label{fig:eff_R_3S1_n1_32}
}
\end{figure}

While the present results are not as satisfactory as for the (6.1 fm)$^3$ box case, 
we find it encouraging that the energy shift relative to the two nucleon threshold,  
which is negative for the ground state, is clearly positive 
for the first excited state and 
is much lower than the value expected for relative
momentum of $2\pi/L$, see Fig.~\ref{fig:eff_R_3S1_n1_32}.  
Because of the presence of a positive slope, the estimate of the energy shift suffers 
from a sizable systematic error from the choice of the fit range.   
We estimate it by making three fits over the ranges $t=9$--13, 11--13, or $12$--14, and 
taking the difference from the first one which we use as the central value. 
For the ground state we use $t=9$--$13$ as the central fit range, and shift the 
minimum and maximum time by $\pm 1$.  The systematic error due to the choice 
of the ${\cal O}_1$ or ${\cal O}_r$ source is also taken into account.  The results 
are summarized in Tables~\ref{tab:DdE} and ~\ref{tab:DdE_n1} for the ground and first excited states, respectively, on the (4.1 fm)$^3$ box for both the $^3$S$_1$ and $^1$S$_0$ channels.

\begin{figure}[!t]
\includegraphics*[angle=0,width=0.6\textwidth]{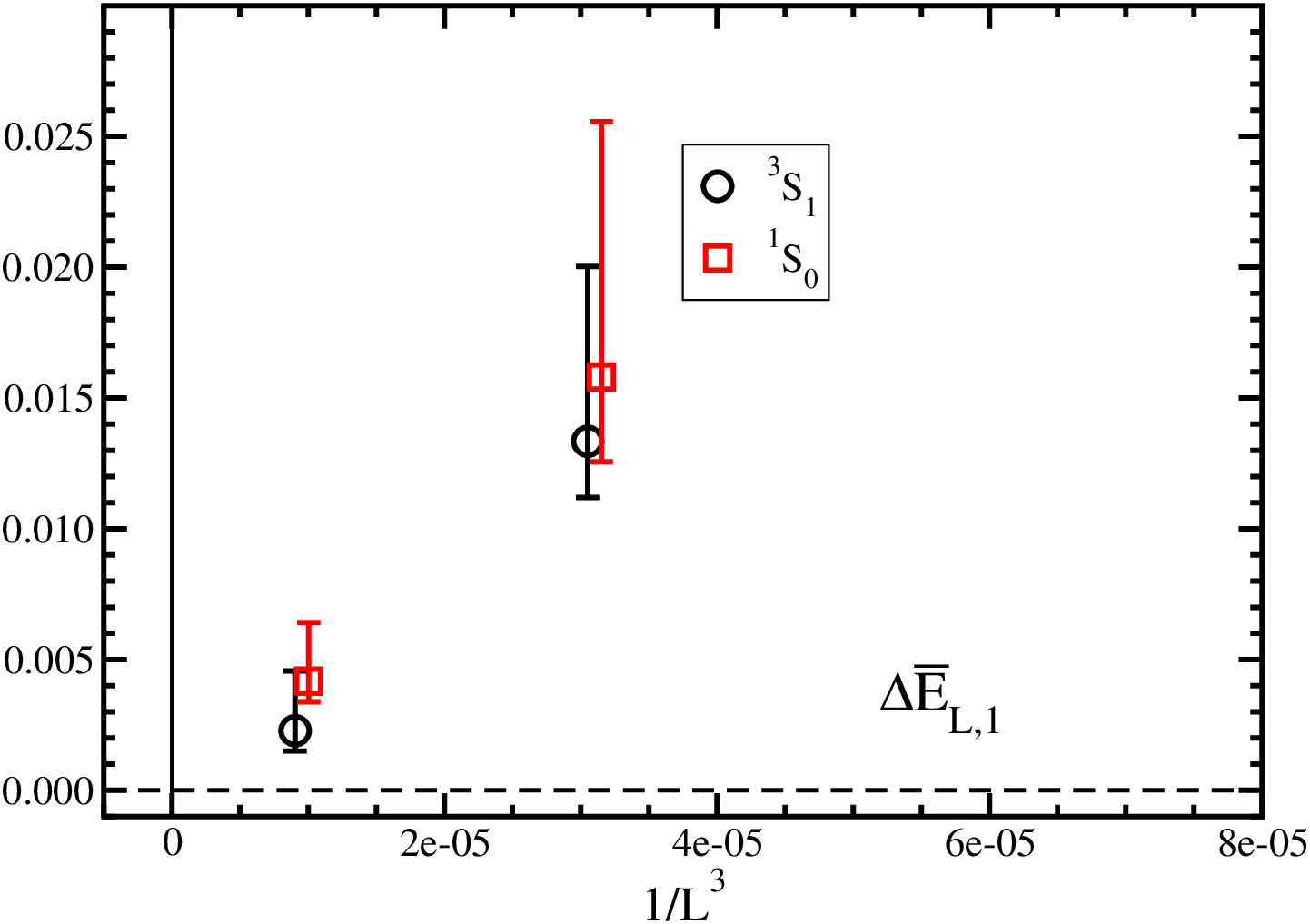}
\caption{
Spatial volume dependence of $\Delta \overline{E}_{L,1}$ in GeV units
for the first excited states in $^3$S$_1$ (circle) 
and $^1$S$_0$ (square) channel.
The square symbols are slightly shifted to positive direction 
in horizontal axis for clarity.
Statistical and systematic errors are added in quadrature.
\label{fig:dE_n1}
}
\end{figure}

In Fig.~\ref{fig:dE_n1} we plot the energy shift for the first excited state from the two lattice 
volumes as a function of $1/L^3$.  A roughly linear behavior, 
with a larger shift  on the (4.1 fm)$^3$ box compared to a smaller shift on the
 (6.1 fm)$^3$ box,  is consistent with this state being a scattering state. 
We evaluate the scattering length using L\"uscher's finite volume formula~\cite{Luscher:1986pf,Luscher:1990ux}, and list them in Table~\ref{tab:DdE_n1},
where we find reasonable consistency between the two volumes.  
If our finding of a bound state in quenched QCD at heavy 
quark mass smoothly continues to the physical point, then this is the first 
calculation which explained a negative scattering length for the deuteron channel.

\subsection{Binding energy from the two calculations}

\begin{figure}[!t]
\includegraphics*[angle=0,width=0.6\textwidth]{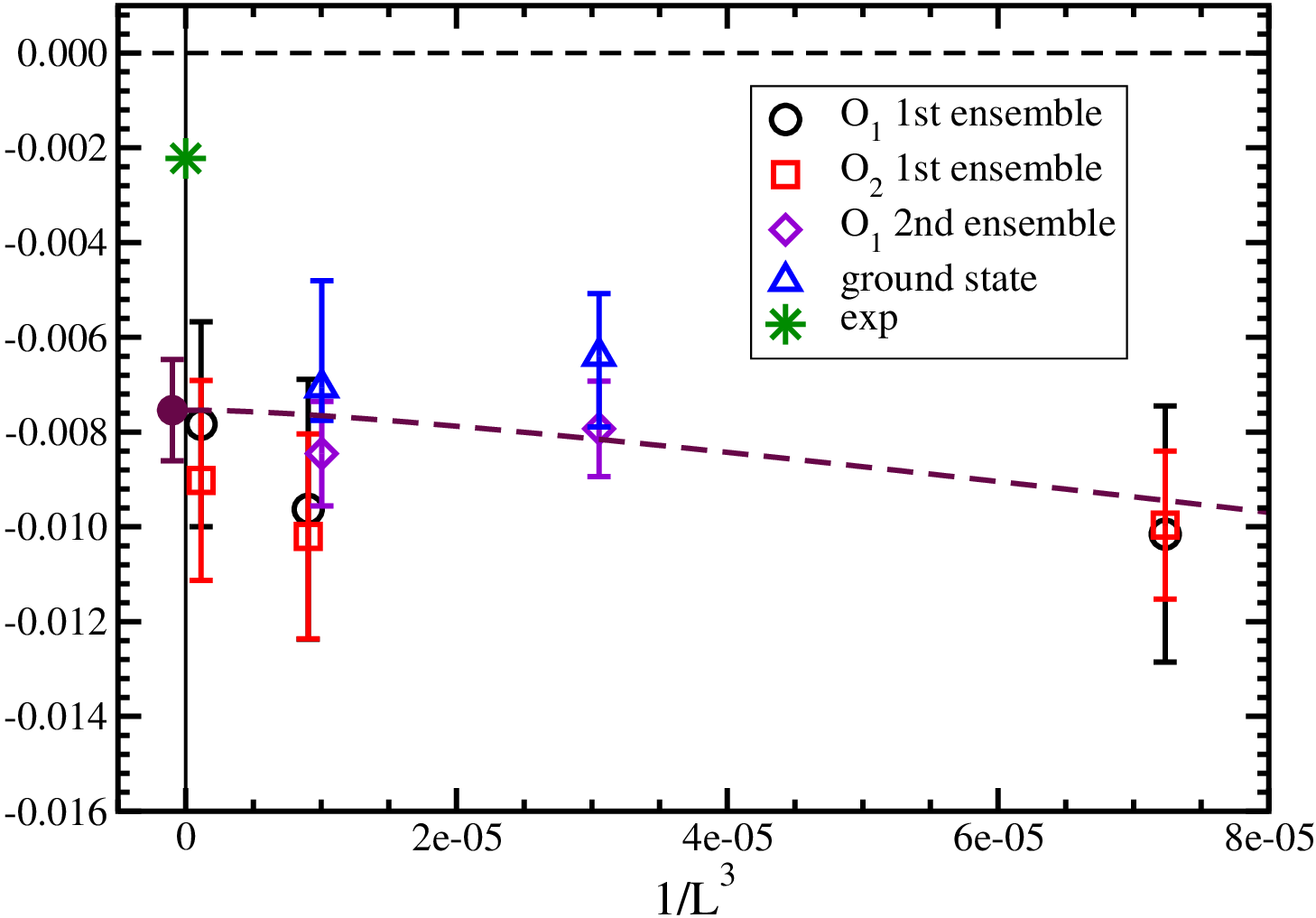}
\caption{
Spatial volume dependence of $\Delta E_L$ and $\Delta \overline{E}_{L,0}$
in GeV units
for $^3$S$_1$ channel with ${\cal O}_1$ (circle), 
${\cal O}_2$ (square) sources on the
first ensembles and $O_1$ source on the second ensemble (diamond).
Result of $\Delta \overline{E}_{L,0}$ obtained by 
diagonalization method is denoted by triangle symbol.
The diamond and triangle symbols at $1/L^3 \approx 10^{-5}$
are slightly shifted to positive direction 
in horizontal axis for clarity.
Statistical and systematic errors are added in quadrature.
Extrapolated results to the infinite spatial 
volume limit (filled circle) 
and experimental values (star) are also presented.
\label{fig:dE_3S1}
}
\end{figure}

\begin{figure}[!t]
\includegraphics*[angle=0,width=0.6\textwidth]{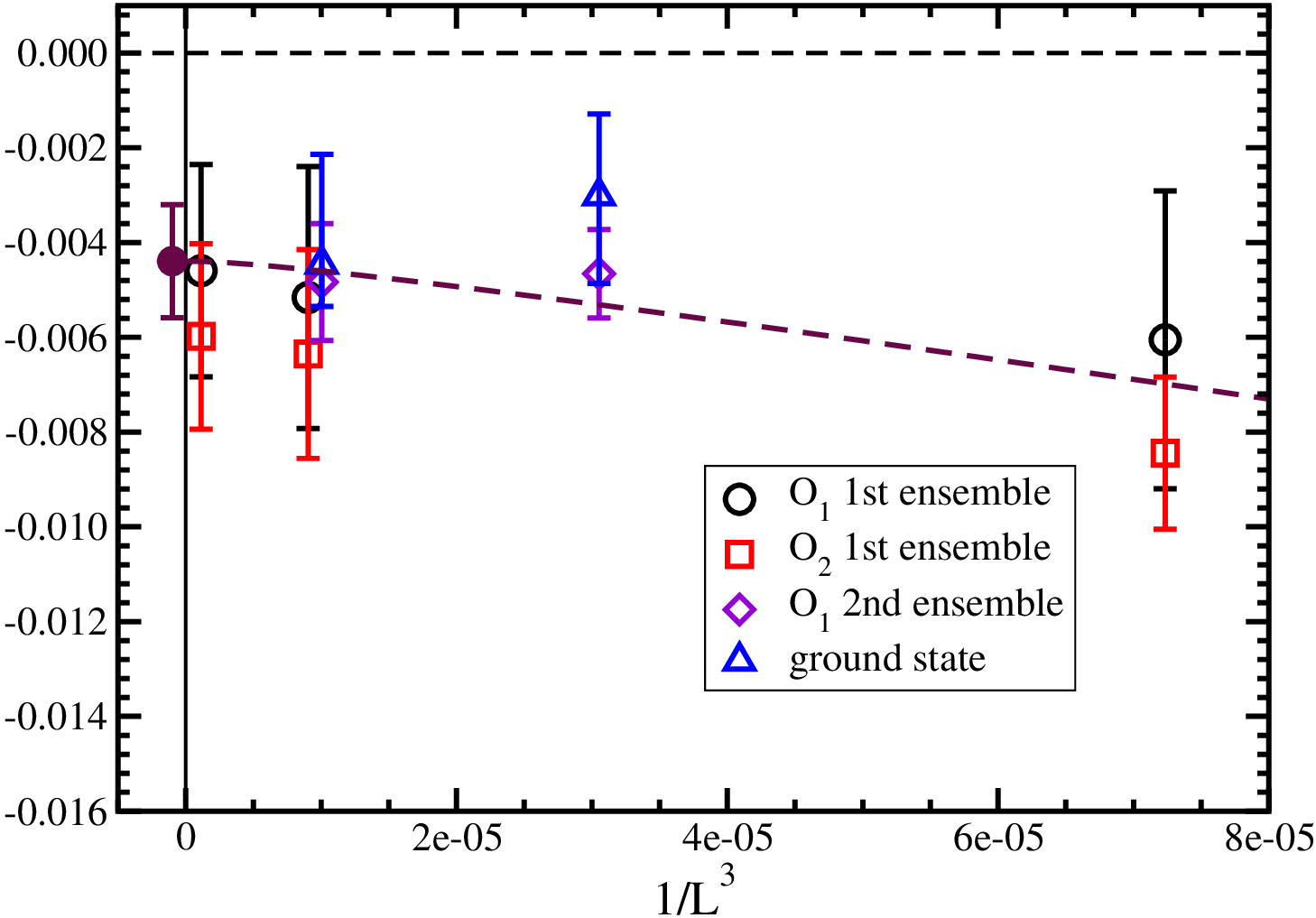}
\caption{
Same as Fig.~\ref{fig:dE_3S1} for $^1$S$_0$ channel.
\label{fig:dE_1S0}
}
\end{figure}

We evaluate the binding energy of the bound state
in the $^3$S$_1$ and $^1$S$_0$ channels using 
the combined results obtained from both the first and second calculations.
Figures~\ref{fig:dE_3S1} and \ref{fig:dE_1S0} are 
the same as Figs.~\ref{fig:dE_3S1_wo} and \ref{fig:dE_1S0_wo}, respectively,
but including the results of the second calculations. 
The new data are reasonably consistent with the previous ones.
We apply the same extrapolation procedure to the infinite volume limit 
as in Sec.~\ref{sec:1st}.
From the fits we obtain the following binding energy for the two channels:
\begin{equation}
-\Delta E_{\infty} = \left\{
\begin{array}{ccl}
7.5(0.5)(0.9) & \mathrm{MeV} & \mathrm{for}\ ^3\mathrm{S}_1,\\
4.4(0.6)(1.0) & \mathrm{MeV} & \mathrm{for}\ ^1\mathrm{S}_0,\\
\end{array}
\right.
\end{equation}
where the first and second errors are statistical and systematic.
These results are reasonably consistent
with the ones in Sec.~\ref{sec:1st}.

\section{Conclusion and discussion}
\label{sec:summary}

We have carried out two calculations in quenched QCD to investigate
whether the two nucleon systems
are bound or not at the heavier quark mass, corresponding
to $m_\pi = 0.8$ GeV.
In the first calculation, we have focused on the ground state
of the two-nucleon system, and have investigated the volume dependence 
of the energy shifts obtained with two different source operators.
We have found that the ground state in the $^3$S$_1$ channel
has little volume dependence,
and a finite energy shift remains in the infinite volume limit.
Based on these results we have concluded that the ground state is
a bound state at the heavy quark mass.
A similar result is obtained in the $^1$S$_0$ channel, though 
the binding energy is almost half of the one in the $^3$S$_1$ channel.

In the second calculation we have carried out two-state analysis using
the diagonalization method. The ground and first excited states are well
separated on the (6.1 fm)$^3$ box, and the ground state energies
for the two  channels
agree with the ones obtained from the single state analysis.
The energy of the first excited state is positive and far below
the free two-nucleon energy with the lowest relative momentum
in both channels.
This leads to the conclusion that each channel has 
one bound state.
We obtain similar results on the (4.1 fm)$^3$ box, though
the contaminations from higher excited states may be larger than
the (6.1 fm)$^3$ case.
The energy of the first
excited state increases as the volume diminishes.
The scattering length is obtained from the energy of the first excited state
using the finite volume formula.
The results in the two volumes reasonably agree with each other.  
In the $^3$S$_1$ channel  the scattering length 
is roughly one fifth of the experimental value.
The difference might be attributed to the heavier quark mass
employed in this calculation.

The existence of the bound state and the negative scattering length 
in the $^1$S$_0$ channel looks odd from the experimental point of view.
In addition we cannot directly compare our result with
those of the model calculations, which
are restricted around physical quark masses.
We expect that the bound state vanishes at some lighter quark mass, 
where the scattering length diverges changing the sign 
from negative to positive. 
Further reduction of the quark mass would decrease the scattering length.
Confirmation of this scenario requires to investigate the quark mass
dependences of the binding energy and the scattering length.
We leave this study to future work.

\section*{Acknowledgments}
Numerical calculations for the present work have been carried out
on the HA8000 cluster system at Information Technology Center
of the University of Tokyo, on the PACS-CS computer 
under the ``Interdisciplinary Computational Science Program'' of 
Center for Computational Sciences, University of Tsukuba, 
and on the T2K-Tsukuba cluster system at University of Tsukuba. 
We thank our colleagues in the PACS-CS Collaboration for helpful
discussions and providing us the code used in this work.
This work is supported in part by Grants-in-Aid for Scientific Research
from the Ministry of Education, Culture, Sports, Science and Technology 
(Nos. 18104005, 18540250, 22244018) and 
Grants-in-Aid of the Japanese Ministry for Scientific Research on Innovative 
Areas (Nos. 20105002, 21105501, 23105708).

\bibliography{paperNN}

\clearpage
%
%
%
%
\begin{table}[!t]
\caption{
Number of configurations ($N_{\rm conf}$), 
number of measurements on each configuration ($N_{\rm meas}$),
acceptance rate in the HMC algorithm,
pion mass ($m_\pi$) and nucleon mass ($m_N$) for the first ensembles.
\label{tab:conf_meas}
}
\begin{ruledtabular}
\begin{tabular}{cccccc}
$L$ & $N_{\mathrm{conf}}$ & $N_{\mathrm{meas}}$ 
& accept.(\%)
& $m_\pi$ [GeV] & $m_N$ [GeV]\\
\hline
24 & 2500 &  2 & 93 & 0.8000(3) & 1.619(2) \\
48 &  400 & 12 & 93 & 0.7999(4) & 1.617(2) \\
96 &  200 & 12 & 68 & 0.8002(3) & 1.617(2) \\
\end{tabular}
\end{ruledtabular}
\end{table}
\begin{table}[!t]
\caption{
\label{tab:dE}
Energy shift $-\Delta E_L$ in MeV units for $^3$S$_1$ and $^1$S$_0$ channels 
on each spatial volume with the first ensembles. 
Extrapolated results to the infinite spatial volume limit 
are also presented. The first and second errors are
statistical and systematic, respectively.
}
\begin{ruledtabular}
\begin{tabular}{ccccc}
& \multicolumn{2}{c}{$^3$S$_1$}&
\multicolumn{2}{c}{$^1$S$_0$}\\
$L$  & ${\cal O}_1$ & ${\cal O}_2$ & ${\cal O}_1$ & ${\cal O}_2$ \\
\hline
24 & 10.2(2.2)(1.6) & 10.0(1.5)(0.5) & 6.1(2.3)(2.2) & 8.4(1.5)(0.5)\\
48 &  9.6(2.6)(0.9) & 10.2(2.0)(0.8) & 5.2(2.6)(0.8) & 6.4(2.0)(0.8)\\
96 &  7.8(2.1)(0.4) &  9.0(2.0)(0.5) & 4.6(2.0)(1.1) & 6.0(1.9)(0.5)\\
$\infty$ & \multicolumn{2}{c}{9.1(1.1)(0.5)} &
\multicolumn{2}{c}{5.5(1.1)(1.0)}\\
\end{tabular}
\end{ruledtabular}
\end{table}
\begin{table}[!t]
\caption{
Number of configurations ($N_{\rm conf}$), 
separation of trajectories between each measurement ($N_{\rm sep}$),
number of measurements with ${\cal O}_1$ 
source on each configuration ($N_{\rm meas}$),
number of $Z(3)$ random number for $O_r$ source 
on each configuration ($N_{\rm rand}$),
spatial interval between smeared quark fields for ${\cal O}_r$ 
source ($N_{\rm mod}$),
acceptance rate in the HMC algorithm,
pion mass ($m_\pi$) and nucleon mass ($m_N$)
for the second ensembles.
\label{tab:conf_meas_2nd}
}
\begin{ruledtabular}
\begin{tabular}{ccccccccc}
$L$ & $N_{\mathrm{conf}}$ & $N_{\mathrm{sep}}$ 
& $N_{\mathrm{meas}}$ & $N_{\mathrm{rand}}$ 
& $N_{\mathrm{mod}}$ & accept.(\%)
& $m_\pi$ [GeV] & $m_N$ [GeV]\\
\hline
32 &  300 & 400 & 192 & 40 & 16 & 87 & 0.7998(2) & 1.6162(9) \\
48 &  300 & 200 & 144 & 32 & 12 & 93 & 0.8001(1) & 1.6176(4) \\
\end{tabular}
\end{ruledtabular}
\end{table}
\begin{table}[!t]
\caption{
\label{tab:DdE}
Energy shifts $-\Delta E_L$ and $-\Delta \overline{E}_{L,0}$ in MeV units
for $^3$S$_1$ and $^1$S$_0$ channels at $L=32$ and 48 on the second ensembles. 
The first and second errors are
statistical and systematic, respectively.
}
\begin{ruledtabular}
\begin{tabular}{ccccc}
& \multicolumn{2}{c}{$^3$S$_1$}&
\multicolumn{2}{c}{$^1$S$_0$}\\
$L$ & {$-\Delta E_L$} & {$-\Delta\overline{E}_{L,0}$} & 
{$-\Delta E_L$} & {$-\Delta\overline{E}_{L,0}$}
\\
\hline
32 & 7.9(0.6)(0.8) & 6.4(1.3)$\left(^{+0.7}_{-0.1}\right)$
   & 4.7(0.7)(0.6) & 3.0(1.7)$\left(^{+0.7}_{-0.3}\right)$\\
48 & 8.5(1.1)(0.3) & 7.1(0.7)$\left(^{+0.1}_{-2.2}\right)$
   & 4.8(1.0)(0.7) & 4.5(0.9)$\left(^{+0.1}_{-2.1}\right)$\\
\end{tabular}
\end{ruledtabular}
\end{table}
\begin{table}[!t]
\caption{
\label{tab:DdE_n1}
Energy shift of the first excited state $\Delta \overline{E}_{L,1}$ 
and scattering length $a_0$
for $^3$S$_1$ and $^1$S$_0$ channels at $L=32$ and 48
after diagonalization in two-state analysis.
The first and second errors are
statistical and systematic, respectively.
}
\begin{ruledtabular}
\begin{tabular}{ccccc}
& \multicolumn{2}{c}{$^3$S$_1$} & \multicolumn{2}{c}{$^1$S$_0$}\\
$L$ & $\Delta \overline{E}_{L,1}$ [MeV] & $a_0$ [fm]
& $\Delta \overline{E}_{L,1}$ [MeV] & $a_0$ [fm]\\
\hline
32 & 13.3(1.3)$\left(^{+6.6}_{-1.7}\right)$ 
& $-1.5(0.2)\left(^{+0.2}_{-1.4}\right)$
& 15.8(1.6)$\left(^{+9.6}_{-0.3}\right)$ 
& $-1.8(0.3)\left(^{+0.4}_{-12.9}\right)$\\
48 & 2.3(0.8)$\left(^{+2.2}_{-0.1}\right)$ 
& $-1.05(24)\left(^{+0.05}_{-0.65}\right)$
& 4.2(0.8)$\left(^{+2.1}_{-0.0}\right)$
& $-1.62(24)\left(^{+0.01}_{-0.75}\right)$\\
\end{tabular}
\end{ruledtabular}
\end{table}

\clearpage
%
%
%
%
\clearpage
\clearpage

\end{document}